\documentclass[structabstract]{aa}  
\usepackage{graphicx}
\usepackage{txfonts}
\begin{document}
\title{The fate of heavy elements in dwarf galaxies -- the role of
  mass and geometry}

   \author{S. Recchi\inst{1}\thanks{simone.recchi@univie.ac.at}\and 
           G. Hensler\inst{1}\thanks{gerhard.hensler@univie.ac.at}}

   \institute{Institute for Astrophysics, University of Vienna,
              T\"urkenschanzstrasse 17, A-1180 Vienna}

   \date{}

 
  \abstract
  {Energetic feedback from Supernovae and stellar winds can drive
    galactic winds.  Dwarf galaxies, due to their shallower potential
    wells, are assumed to be more vulnerable to this phenomenon.
    Metal loss through galactic winds is also commonly invoked to
    explain the low metal content of dwarf galaxies.}
  {Our main aim in this paper is to show that galactic mass cannot be
    the only parameter determining the fraction of metals lost by a
    galaxy.  In particular, the distribution of gas must play an
    equally important role.}
  {We perform 2-D chemo-dynamical simulations of galaxies
    characterized by different gas distributions, masses and gas
    fractions.}
  {The gas distribution can change the fraction of lost metals through
    galactic winds by up to one order of magnitude.  In particular,
    disk-like galaxies tend to loose metals more easily than roundish
    ones.  Consequently, also the final metallicities attained by
    models with the same mass but with different gas distributions can
    vary by up to one dex.  Confirming previous studies, we also show
    that the fate of gas and freshly produced metals strongly depends
    on the mass of the galaxy.  Smaller galaxies (with shallower
    potential wells) more easily develop large-scale outflows,
    therefore the fraction of lost metals tends to be higher.}
  {}


   \maketitle
%

\section{Introduction}

Theories of cold dark matter-dominated hierarchical growth of
structures predict that dwarf galaxy- (DG-)sized objects are the
building blocks for the formation of large galaxies.  In spite of
their relevance, the most important physical phenomena regulating the
birth and evolution of DGs are still obscure to date.  For sure, star
formation (SF) plays a key role in shaping DGs and determining their
fates.  Since the binding energy of the interstellar medium (ISM) in
early gas-rich DGs is very small (smaller than the explosion energy of
just a few Supernovae (SNe)) many authors speculated that a high SF
rate in a DG would create a galactic wind and thus produce a
transition from a dwarf irregular (dIrr) to a dwarf spheroidal (dSph)
or dwarf elliptical (dE) (see e.g. Larson \cite{lars74}; Vader
\cite{vader86}; Dekel \& Silk \cite{ds86}).  From a chemical point of
view, the occurrence of a galactic wind right after the formation of
the first stars would imply a very limited interval of time during
which the metals, restored by dying stars, can pollute the ISM and
enrich the following stellar populations.  If galactic winds
preferentially occur in DGs, these objects (with low masses) will
experience a limited chemical enrichment.  Therefore, a correlation
between stellar mass M (or luminosity L) and metallicity Z of DGs is
expected.  Indeed, this M-Z correlation among DGs exists (see e.g.
Skillman et al.  \cite{skh89}; Lee et al.  \cite{lee06}; Kirby et al.
\cite{kirby08}; Zhao et al. \cite{zhao10}; Andrews \& Martini
\cite{am12}).  A M-Z relation extends also to high redshifts (Erb et
al.  \cite{erb06}; Maiolino et al.  \cite{maio08}; Laskar et al.
\cite{lask11}; Wuyts et al.  \cite{wuyts12}) although, in this case,
the galactic masses of the galaxies for which metallicity
determinations are available are usually quite high (but see Mannucci
et al.  \cite{mann11}).  The M-Z relation among DGs corroborates the
idea that SN-driven galactic winds play a dominant role in the
evolution of these objects.  More recently, a general relation between
stellar mass, gas-phase metallicity and star formation rate has been
found (Mannucci et al. \cite{mann10}).

\begin{figure}
   \centering
   \includegraphics[width=10cm]{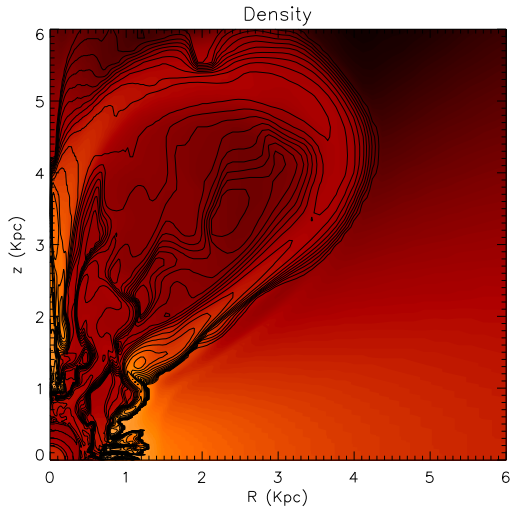}
   \caption{Gas (filled contours) and oxygen (black contour lines)
     distribution of a model galaxy (taken from Recchi et al.
     \cite{recc06}) experiencing a galactic outflow.  Brighter colors
     indicate larger gas densities.  }
         \label{fig:gas_O}
\end{figure}

However, detailed hydrodynamical simulations of DGs showed that the
galactic winds, although often able to expel a large fraction of
freshly produced metals, are unable to eject an equally large fraction
of pristine (i.e. not processed) gas.  This is mostly due to the fact
that, if the initial DG gas distribution is flattened (as observed in
dIrrs), then there is a direction with steeper pressure gradient so
that the galactic wind will preferentially expand along that
direction, the transport of gas along the other directions being very
limited (see e.g. D'Ercole \& Brighenti \cite{db99}; MacLow \& Ferrara
\cite{mf99} (hereafter MF99); Recchi et al. \cite{rmd01}).  This
effect can be appreciated by inspecting Fig. \ref{fig:gas_O}, where
the gas and oxygen distribution of a model galaxy experiencing a
galactic wind (taken from the calculations of Recchi et al.
\cite{recc06}) is shown.  Most of the disk gas (at least above R=1
kpc) has not been affected by the galactic wind and it will fall back
towards the center of the galaxy once the energy source of the
starburst will be exhausted.  On the other hand, the oxygen can be
easily channelled along the funnel created by the galactic wind.  Also
very energetic ($E \sim 10^{53}$ erg) hypernovae exploding in dwarf
protogalaxies, although more disruptive than normal SNe, are able to
expel only $\sim$ 10 to 20 \% of the baryonic mass originally present
in the galaxy (Vasiliev et al.  \cite{vvs08}).

This result is widely accepted by the astrophysical community,
although some authors (e.g. Tenorio-Tagle \cite{tt96}; Silich \&
Tenorio-Tagle \cite{stt98}), believe that low-density gaseous galactic
halos can strongly affect the circulation of the metal-rich matter
processed by the central starburst.  Kiloparsec-scale bipolar outflows
are created but, later on, loss of pressure support and interaction
with the diffuse halo slow down the expansion of the superbubble and
the freshly produced metals can eventually fall back towards the
center of the galaxy.  Thus, according to this scenario, energetic
events associated to a starburst can still create a large-scale
outflow but this outflow does not necessarily become a galactic wind
(i.e. its velocity does not exceed the escape velocity of the galaxy).

It is also worth noticing that, while the chemical yields of exploding
massive stars are very large (thus the metallicity of interiors of
SN-driven bubbles is up to 40 times the solar value), abundances of
the hot gas in galaxies experiencing large-scale outflows (determined
from X-ray spectra) cover with high probability a range between only
solar and twice solar (see e.g.  Martin et al.  \cite{mart02} for
NGC1569 or Ott et al. \cite{owb05} for a larger sample of galaxies).
From this fact, it can be deduced that the hot SN ejecta becomes mass
loaded (by a factor of up to 10) during its expansion, probably due to
evaporative and turbulent mixing with colder interstellar clouds.  By
this, the hot gas loses not only momentum but also energy by enhanced
cooling in addition to adiabatic expansion so that the outbreak of the
hot gas to a galactic wind could be hampered in many cases.  This
mass-loading effect is probably causing the discrepancy between the
observed extent and temperature of the extremely cool superbubble in
NGC~1705 and its analytic parameter correlation (Hensler et al.
\cite{hens98}).

Because of the gas density outside the SF region, the development of
galactic outflows or galactic winds considerably depends on the degree
of flattening (or on the rotation) of the parent galaxy, an aspect
that has not been fully explored in the past (but see Silich \&
Tenorio-Tagle \cite{stt01}; Michielsen et al. \cite{mich07}; Schroyen
et al.  \cite{schr11}. See also Strickland \& Stevens \cite{ss00};
Ferrara \& Tolstoy \cite{ft00}; Vasiliev et al.  \cite{vvs08}; Recchi
et al.  \cite{rha09}).  Although flat galaxies are supposed to loose a
large fraction of metals as a consequence of bipolar galactic winds,
roundish galaxies are not characterized by a direction along which the
pressure gradient is significantly steeper, therefore they are more
likely to retain much of the metals (Marcolini et al. \cite{marc06};
Recchi et al.  \cite{recc07}).  In the end, this difference in the
initial gas distribution could lead to a spread in the observed final
metallicity of galaxies with the same mass.  Such a spread is indeed
observed in the M-Z relation of DGs (Lee et al.  \cite{lee06}; Kirby
et al.  \cite{kirby08}; Zhao et al. \cite{zhao10}), or in the same
correlation at $z\simeq 3.5$ (Maiolino et al.  \cite{maio08}).  In
this paper we explore in detail the effect of gas geometry on the
development of galactic winds and on the fate of freshly produced
metals.  This work must be thus seen as a refined study along the
lines of the MF99 work, in which variations in the initial
distribution of gas in galaxies have not been considered.  The second
key parameters we consider in this study is the initial mass of the
galaxy.  As already mentioned, small galaxies (with shallow potential
wells), are expected to develop galactic winds more easily than larger
galaxies, therefore the mass is certainly a key parameter in
determining the fate of pristine gas and freshly produced heavy
elements.

The dynamical evolution of a galaxy is so complex that the fate of
heavy elements cannot depend only on mass and geometry of the parent
galaxy.  An obvious parameter that strongly affects the development of
galactic winds is the luminosity (MF99).  In turn, the luminosity
depends on the star formation rate (SFR) and on the star formation
history (SFH) and it is known, at least in the Local Group, that very
large galaxy-to-galaxy SFH variations exist (see e.g. Monelli et al.
\cite{mone10_1}, \cite{mone10_2}; Hidalgo et al. \cite{hida11}).
Moreover, other details of the ISM structure (for instance the
presence of clouds or porosity) and of the feedback prescriptions
(e.g. how the SN explosion energy is redistributed, where, on which
timescales and at which efficiency etc.) can all play a role in
determining the amount of metals carried out of a galaxy by galactic
winds and eventually in the final global metallicity of the galaxy.
Although the consequences of an inhomogeneous ISM on the development
of galactic winds is analyzed in detail in Recchi \& Hensler
(\cite{rh07}, hereafter RH07), we have undertaken a comprehensive
parameter study through detailed chemo-dynamical simulations, the
results of which will be presented elsewhere.  Here we will focus
mainly on 18 basic models, characterized by different geometries,
total baryonic masses and initial gas fractions.  The paper is
organized as follows: in Sect. \ref{sec:summ} a short summary on the
literature results concerning the fate of freshly produced metals will
be given.  In Sect. \ref{sec:setup} our numerical scheme will be
briefly reviewed and the set-up of the 18 basic models models
described.  In Sect.  \ref{sec:results} the main results of these
calculations will be presented, underlying also the effect of some
other key parameters not fully explored in this work.  In Sect.
\ref{sec:disc} the results will be discussed and some conclusions will
be drawn.

\section{A brief overview of literature results}
\label{sec:summ}
As already mentioned in the Introduction, several papers in the
literature attempted to study the effect of galactic winds on the
circulation and redistribution of metals in DGs.  The main results of
the often-cited work of MF99 are that, even in the presence of a
strong galactic wind driven by SNeII, the ejection efficiency of
unprocessed gas is always close to zero (with the exception of
galaxies with initial baryonic mass $\leq$ 10$^6$ M$_\odot$).  On the
other hand, the ejection efficiency of freshly produced heavy elements
is always close to one (with the exception of galaxies with initial
baryonic mass $\geq$ 10$^9$ M$_\odot$).  D'Ercole \& Brighenti
(\cite{db99}) found results similar to the ones of MF99, in the sense
that only minor fractions of ISM (but in some cases large fractions of
metals) are expelled through galactic winds.  They however calculated
the evolution of the galaxy for much longer times than MF99,
discovering that the center of the galaxy can be replenished with cold
gas in a timescale of $\sim$ 100 Myr (see also Recchi \& Hensler
\cite{rh06}).  As already mentioned, Silich \& Tenorio-Tagle
(\cite{stt98}) found instead that in most of the models galactic winds
do not develop (mainly due to the presence of a hot gaseous halo
surrounding the galaxy).  Fragile et al. (\cite{frag04}) studied the
effect of SN explosions not localized in the center of the galaxy (as
MF99 did) but distributed over radii of up to 80\% of the disk radius,
discovering that, in this case, radiative losses are more effective
and the development of a large-scale outflow is hampered.  Also very
interesting is the work of Scannapieco \& Br\"uggen (\cite{sb10}), who
attempted to model turbulent velocities and turbulent length scales in
DGs and to inject SN energy into supersonic turbulence.  The wind
efficiencies they found are still quite low.
Rodr{\'{\i}}guez-Gonz{\'a}lez et al.  (\cite{rodr11}) addressed the
same problem (mass and metal ejection efficiencies in DGs) exploring a
much more extended set of parameters (in particular they covered a
wide range of starburst and galactic gas masses).  Their results are
overall in agreement with the previously quoted works (with the
exception of Silich \& Tenorio-Tagle \cite{stt98}).  More recently,
Schroyen et al. (\cite{schr11}) explored in detail the effect of
rotation on the evolution of DGs.  Their focus was to understand the
origin of the dichotomy in radial stellar metallicity profiles of DGs
and their ``centrifugal barrier mechanism'' help explaining these
observations.  Usually their models do not develop galactic winds (or
develop only weak outflows), thus the determination of ejection
efficiencies (of pristine gas and freshly produced metals) is not
addressed in their paper.  The same can be said about the detailed
determination of the energy required to expel newly processed matter
by Silich \& Tenorio-Tagle (\cite{stt01}); their analytic model is in
good agreement with the numerical results of MF99 and shows clearly
the effect of geometry on the development of galactic winds and on the
fate of metals if the intergalactic pressure is low.  However, their
inferred large intergalactic pressure hampers the development of
galactic winds.

As this short summary indicates, the past literature mostly focused on
the effect of mass and luminosity on the gas and metal ejection
efficiencies from galaxies.  The chemical evolution is usually not
very detailed in these models.  Recchi et al. (\cite{rmd01},
\cite{recc02}, \cite{recc04}, \cite{recc06}) provided detailed
information on the ejection efficiencies of single chemical species
produced by SNe (of both Type II and Type Ia) and by winds from
intermediate-mass stars.  RH07 added important information on the
effect of clouds.  The most significant results of these works can be
summarized as follows:

\begin{itemize}

\item Galactic-scale outflows carry out of the galaxy mostly the
  chemical elements produced by dying stars during the most recent
  episodes of SF, with large escape fraction of metals with delayed
  production (like Fe and N).

\item Models with very short burst(s) of SF can cool and mix the newly
  formed metals in a very short timescale, whereas, when the SFH is
  continuous and long-lasting, most of the metals are either directly
  ejected outside the galaxy through galactic winds or are confined in
  a hot medium, therefore cannot contribute to the chemical enrichment
  of the warm ionized medium observed by emission lines from the HII
  gas.

\item Models with complex and long-lasting SF episodes taken from the
  literature reproduce the chemical composition and the abundance
  ratios, observed in galaxies, much better than models with bursting
  SF.

\item Clouds (if present) are able to increase the main density of the
  cavity created by the ongoing SF activity, provoking a reduction of
  the total thermal energy by $\sim$ 20 -- 40\% compared to models
  without clouds.

\item The interaction clouds-supershell leads to strong structuring
  and piercing of the shell, allowing the venting out of metals in
  spite of the reduced thermal energy.  The development of large-scale
  outflows is therefore generally delayed but the ejection efficiency
  of metals remains unchanged.

\item Since the clouds are assumed to have a low metallicity, their
  mixing with the ISM tends to reduces the abundance of heavy elements
  of the galaxies.  The models with infalling clouds have thus
  metallicities up to $\sim$ 0.4 dex lower than corresponding models
  without clouds.  The abundance ratios remain unaltered.
\end{itemize}

A study of the early evolution of tidal DGs (which are assumed to be
dark-matter- (DM-)free) and their possible correlation with the
satellite DGs surrounding the Milky Way has been also performed
(Recchi et al. \cite{recc07}).  These models, in spite of being
DM-free, can sustain the energy released by dying stars without
experiencing a complete blow-away, for several hundreds of Myr,
provided that they keep their initial spherical symmetry.  A typical
model develops through a network of cavities and filaments due to the
patchy distribution of the SF sites. A supershell is formed, but it
grows slowly in size and does not quench the SF process.  This is
because the initial ISM distribution in the galaxy is spherical and
there is therefore no preferential direction through which the
galactic wind can flow.  Under these conditions, either the feedback
is able to remove all the gas at once ({\it blow-away}) or all the gas
(or most of it) is retained inside the galaxy.  Our study, and other
similar ones (e.g. Marcolini et al. \cite{marc06}), show that it is
not easy to get rid of all the gas in an initially spherical galaxy.

\section{Description of the model}
\label{sec:setup}

This work aims at studying in detail the dynamical and chemical
evolution of DGs and the effect of key parameters (in particular
baryonic mass, initial gas mass fraction and degree of flattening) on
the development of galactic winds and on the fate of freshly produced
metals.  A follow-up study will extend the parameter space and will
concentrate on the details of the SFH and on the effect of boundary
conditions.  The ultimate goal is to understand the main mechanisms
determining the metallicity and the abundance ratios in DGs and, at
the same time, infer the degree of metal pollution of the
inter-galactic and intra-cluster medium due to these objects.

\subsection{The chemo-dynamical code}
\label{subs:model}

As chemo-dynamical code we use basically the one of Recchi et al.
(\cite{recc07}).  We recall here the basic features of this code.  It
is a 2-D code in cylindrical coordinates based on a second-order,
MUSCL-type upwind scheme (the 1-D version of this scheme is described
in Bedogni \& D'Ercole \cite{bd86}).  The chemical enrichment of the
galaxy is followed in detail; the production (by SNe of Type II, Type
Ia and intermediate-mass stars) of 8 chemical elements (H, He, C, N,
O, Mg, Si and Fe) is considered and the advection of these elements is
followed by means of passive scalar fields.  For instantaneous bursts,
this scheme is described in detail in Recchi et al. (\cite{rmd01}) and
its extension to more complex SFHs is explained in Recchi et al.
(\cite{recc04}).  Since the code keeps correctly track of the
evolution of the metallicity in each computational cell, the detailed
metallicity-dependent cooling function of B\"ohringer \& Hensler
(\cite{bh89}) can be implemented.  The heat transport equation is also
solved by means of the Crank-Nicolson method (see D'Ercole \&
Brighenti \cite{db99} for details) using the classical Spitzer-H\"arm
thermal conductivity (Spitzer \& H\"arm \cite{sh53}; Spitzer
\cite{spit56}).  A saturated heat flux (Cowie \& McKee \cite{cm77}) is
adopted if the mean free path of electrons is larger than the
temperature scalelength.  Some improvements of this code are described
in Recchi et al. (\cite{recc07}).  In particular,
metallicity-dependent stellar winds from massive and intermediate-mass
stars have been considered and, primary and secondary nucleosynthetic
production by stars of any age and any metallicity has been carefully
calculated.  Recchi et al.  (\cite{recc07}) describe also an
implementation of SF recipes.  The self-consistent SF module is not
adopted in this work where, in analogy with many previous studies
(MF99 for instance) the SFH (or the luminosity) is an input of the
model and not the result of the galactic evolution.  In a work in
preparation, models with a self-consistent SFH will be shown and
discussed.  Moreover, although in Recchi et al. (\cite{recc07}) a
self-gravity solver (based on Rieschick \& Hensler \cite{rh03}) has
been implemented, we decided not to adopt it in this study neither.
The main reason for this choice is that we neglect self-gravity in
building the initial equilibrium configuration (see below in the next
Subsection).  A parallel line of research (Vorobyov et al.
\cite{voro12}; Vorobyov et al., in preparation) analyzes in detail DG
models with a consistent implementation of the gas self-gravity.  The
inclusion of self-gravity in the code would imply that, at the
beginning of the simulation, the gas is initially out of equilibrium
even without sinks or sources of energy.  This would establish an
inward gas flow.  Since our main aim in this paper is to study the
effects of geometry and other parameters on the flow rate of gas and
metals, we do not want gas infall to affect our results and we want
flows of gas (and metals) to be solely affected by feedback processes.
Moreover it is important to notice that most of similar papers already
described in Sect. \ref{sec:summ} (e.g. MF99, D'Ercole \& Brighenti
\cite{db99}, Strickland \& Stevens \cite{ss00}) also neglect
self-gravity.

\subsection{The set-up}
\label{subs:setup}

We start with a reliable set-up of a DG, in which the gas is initially
in isothermal equilibrium with a spherical DM halo and with the
centrifugal force.  Analogously to what is done in many similar
studies, we neglect self-gravity in building the initial equilibrium
configuration.  We outline here that this is in general not correct
even if most of the mass of the model galaxy is in the form of DM and
the correct derivation of an equilibrium configuration is the one
outlined by Vorobyov et al. (\cite{voro12}).  However, the equilibrium
configurations produced by Vorobyov et al.  (\cite{voro12}) are quite
complex (and also computationally demanding) and it is difficult to
obtain two analogous equilibrium configurations, differing only for
the degree of flattening of the initial gas distribution.

We consider three possible values for the initial baryonic mass of the
galaxy (10$^7$, 10$^8$ and 10$^{9}$ M$_\odot$) with a factor of $\sim$
10 more massive DM halos.  The exact factor is deduced by the
correlation between the dark matter-to-baryon ratio $\phi$ and the gas
content as adopted by MF99, namely $\phi=34.7\,M_{g,7}^{-0.29}$, where
$M_{g,7}$ is the gaseous mass of the galaxy in units of 10$^7$
M$_\odot$).  This correlation is adapted from the work of Persic et
al.  (\cite{pss96}), and is based on fitting procedures of observed
rotation curves in local galaxies.  One should warn the reader that it
is not clear whether present-day dark matter-to-baryon ratios also
correctly reflect the initial conditions of the galaxy, but we
nevertheless adopt this correlation in order to facilitate the
comparison with the results of MF99.  The virial radius of the DM
halo is assumed to be:
\begin{equation}
r_{vir}=0.75 \,M_8^{1/3}\,h^{-1}\,\left(\frac{1+z_{gf}}{10}\right)^{-1}\,
{\rm kpc},
\label{eq:rvir}
\end{equation}
(Madau et al. \cite{mfr01}; Mori et al. \cite{mfm02}).  In this
formula, $M_8$ is the halo mass in units of 10$^8$ M$_\odot$ and
$z_{gf}$ is the redshift of galaxy formation.  Our reference value for
$z_{gf}$ is 8.  It is important also to stress that our initial
distributions of gas are not artificially truncated at some cut-off
radii as the models of MF99.  Therefore, without sources or sinks of
energy, our models preserve their initial configurations (at variance
with the models of MF99 which tend to expand).  However, since our
distribution of gas extends until the edges of the computational box,
we need to establish a radius within which we calculate the mass of
the galaxy.  We take this to be half of the virial radius.  Therefore,
galaxies with the same nominal baryonic mass are normalized in such a
way that the total mass within $0.5 \cdot r_{vir}$ is the same,
irrespective of the degree of flattening.  This implies that flat
galaxies have larger central densities than roundish ones (see below
and Table \ref{table:2}).

A small stellar disk is also initially present in the galaxy.  Its
density is such that a Miyamoto-Nagai potential 
\begin{equation}
\psi(R,z)=-\frac{GM_d}{\sqrt{R^2+\left(a+\sqrt{z^2+b^2}\right)^2}},
\label{eq:mn}
\end{equation}
\noindent
(where $M_d$ is the mass of the stellar disk and $R$, $z$ are the
cylindrical radial and vertical coordinates, respectively), is
reproduced.  The ratio $b/a$ between the scale lengths identifying the
Miyamoto-Nagai potential is taken as one of the key parameters that we
vary in our models.  A small $b/a$ corresponds to a flat model (for
$b/a \rightarrow 0$ the potential tends to the razor-thin Kuzmin
model), whereas if $b/a$ is very large, the galaxy tends to be rounder
(for $b/a \rightarrow \infty$ the potential tends to the Plummer's
spherical potential).  

It is expected that an initially flat distribution of gas results in
an easier development of bipolar outflows, since the pressure gradient
along the polar direction is much steeper than that along the galactic
disk.  As already discussed, along the polar direction a significant
fraction of the matter processed in the central SF region can be thus
channelled and eventually lost from the galaxy.  On the other hand, if
the galaxy is initially spherical (or almost spherical), no
preferential propagation direction for the superbubble exists and the
freshly produced metals are more likely to remain confined inside the
galaxy.  One of the aims of this work is to confirm this empirical
assumption and quantitatively determine the fraction of gas and
metal-rich stellar ejecta lost from a DG, as a function of its degree
of flattening.  We consider three representative values for the ratio
$b/a$: $0.2$ (flat models, designated with the letter ``F''); $1$
(medium models or ``M'') and $5$ (roundish models or ``R'').  However,
the mass of this pre-existing stellar disk cannot be established a
priori, therefore as another parameter we vary the ratio between the
mass of the Miyamoto-Nagai stellar disk and the total baryonic mass
initially present in the galaxy.  In particular, we consider two
basics sets of models; for the first set (designated with ``H'') the
initial gas fraction is high (90\% of the total baryonic mass of the
galaxy) and, consequently, the pre-existing stellar disk represents a
small fraction of the mass budget in the galaxy.  The second set of
models (designated with ``L'') is characterized by a much smaller
initial gas fraction (60\% of the baryonic mass).  Therefore, for
instance, the model H7M represents a galaxy with high initial gas
fraction, 10$^7$ M$_\odot$ of initial baryonic mass and a medium
degree of flattening ($b/a=1$).  We consider thus a total of 18 basic
models, but we will discuss the dependence of our results on some key
parameters in Sect.  \ref{subs:parspace}.

Other key parameters to be varied are the duration and intensity of
the SF episode (as already mentioned, the analysis of models with
self-consistent SFHs will be deferred to a future work).  We assume
here for simplicity that the SF is constant (with some pre-defined
intensity $\psi$) for a period of time $\Delta t$.  The two parameters
$\psi$ and $\Delta t$ are allowed to vary, with the constraint that
for models of equal mass the same final amount of stars is produced.
In particular, we constrain the models ``H'' to produce, at the end of
the SF period, a mass of newly formed stars which is twice as much as
the mass of the initial stellar disk (namely the fraction $f_N =M_{*,
  new}/M_d$ is set to be equal to 2).  For the second set of models,
designated with ``L'', we set $f_N=0.5$, namely (since the
pre-existing stellar disk is more massive for this set of models) we
allow newly formed stars to be only 50\% of the pre-existing disk at
the end of the SF period.  The resulting integrated luminosity of our
models is not far from being constant (a constant luminosity has been
assumed by MF99).  Our reference models have $\Delta t =$ 500 Myr and
a SFR $\psi$ such that, after $\Delta t$, the required fraction $f_N$
is obtained.  It is obvious that, keeping the starburst duration
$\Delta t$ constant, a large starburst luminosity (i.e. a large
$\psi$) produces a stronger outflow, thus in turn larger ejection
efficiencies of gas and metals (see MF99).  It is however not obvious
if short, intense starbursts are more effective than long--lasting,
milder SF episodes in getting rid of gas and processed matter.  The
answer to this question will be sought by considering the dependence
of the ejection efficiencies in models with varying $\psi$ and $\Delta
t$ but with constant $\psi \times \Delta t$.  In particular, a few
models have been calculated in which $\Delta t$ is a factor of 10
smaller than the reference value (i.e. $\Delta t=50$ Myr) and,
consequently, the SFR is ten times more intense.  Our reference value
for $\Delta t$ of 500 Myr is justified by observational studies
indicating that the starbursts in DGs are not as short as previously
thought, lasting on average a few hundred Myrs (see e.g. McQuinn et
al. \cite{mcqu10_1}; \cite{mcqu10_2}).

The initial gas metallicity of our reference sets of model galaxies is
zero.  This is because we do not have a self-consistent evolution of
the early phases of our model galaxies and we can only suppose that
the chemical enrichment of gas and stars in this early phase will be
limited (particularly for the models ``H'').  Moreover, we do not
expect it to be a fundamental parameter in determining the fate of
metals, since the only way a different metallicity can affect the
development of a galactic wind is through the cooling function and the
difference between the zero-metal cooling curve and that of a modest
metallicity (Z=10$^{-2}$ Z$_\odot$ say) is negligible.  We have
nevertheless considered also models in which the initial metallicity
of the stars (and gas!) is observationally determined (in particular
we make use of the M-Z relations observationally determined by
Tremonti et al. \cite{trem04} for the local Universe).

As done by Fragile et al. (\cite{frag04}) and
Rodr{\'{\i}}guez-Gonz{\'a}lez et al. (\cite{rodr11}), another
important parameter in our study is the radius over which the feedback
from dying stars is redistributed.  It is to expect that, if the
``feedback radius'' is very small, the density of the (metal-rich)
ejected material will be so high that radiative losses can be very
significant and can substantially increase the probability that the
ejecta remain locked inside the galaxy (see e.g. Tenorio-Tagle et al.
\cite{tt07}).  It is unclear what happens if the feedback radius is
larger although, as already mentioned, the study of Fragile et al.
(\cite{frag04}) suggests a reduction of ejection efficiencies (of both
gas and starburst matter).  Our reference feedback radius is $R_F=$
200 pc but models with $R_F=50$ pc and $R_F=1000$ pc have been also
computed.

We have also considered the possibility (as we did in RH07) that the
initial distribution of gas is characterized by inhomogeneities.
However, for the sake of simplicity, we consider only a random
perturbation of the initial galactic density distribution.  Namely,
once an equilibrium configuration has been obtained, we modify the
density $\rho(i,j)$ in the grid $(i,j)$ in this way:
\begin{equation}\rho(i,j)=\rho(i,j)\cdot
\left[1+\varepsilon R_{[-1,1]}\right],
\label{eq:pert}
\end{equation}
\noindent
where $\varepsilon$ is a small number (corresponding to the largest
possible amplitude of the perturbation) and $R_{[-1,1]}$ is a randomly
generated number in the range $[-1,1]$.  A more thorough (and
self-consistent) treatment of inhomogeneities and a detailed study of
the effect of boundary conditions (gravitational perturbations,
external pressure, infall of clouds) is deferred to a future paper.

   \begin{figure}
   \centering
   \includegraphics[width=\textwidth]{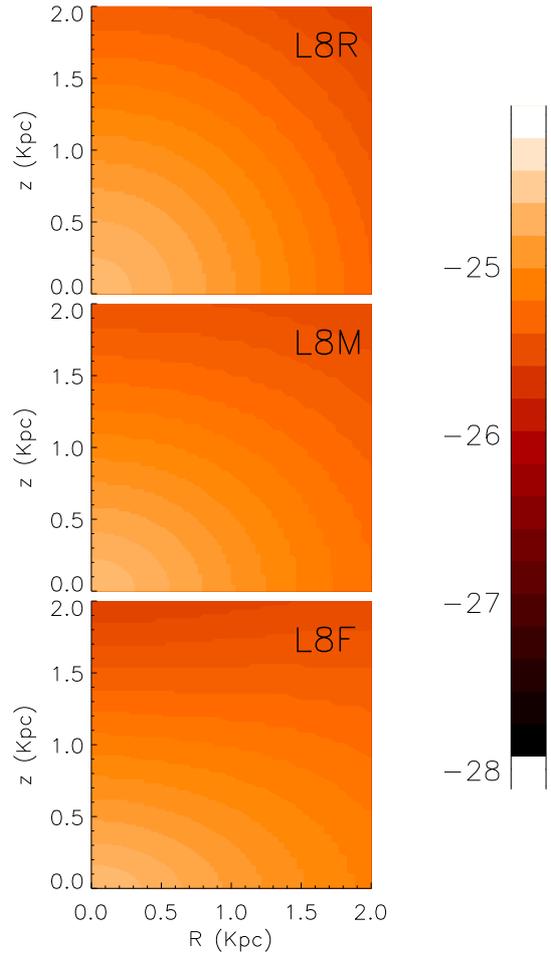}
   \caption{Initial gas distribution for three models (L8R - upper
     panel; L8M - central panel; L8F - lower panel) characterized by
     different degrees of flattening (see Table \ref{table:2}).  The
     density scale (in g cm$^{-3}$) is on the right-hand strip.  }
         \label{fig:init}
   \end{figure}

\begin{table}
\caption{Reference parameter values (common to all 18 basic models)}
\label{table:1}      
\centering                          
\begin{tabular}{c c c c}       
\hline\hline                 
$Z_{in}$$^{(a)}$ & $R_F$$^{(b)}$ & $\Delta t$$^{(c)}$ & $z_{gf}$$^{(d)}$\\    
\hline                        
   0 & 200 & 500 & 8\\      
\hline                                   
\end{tabular}
\tablefoot{$^{(a)}$: Initial gas metallicity;
  $^{(b)}$: Feedback radius, in parsec;
  $^{(c)}$: Duration of the star formation period, in Myr;
  $^{(d)}$: Redshift of galaxy formation}
\end{table}

The central resolution of the simulations is 4 pc for the models with
10$^8$ and 10$^9$ M$_\odot$ of initial baryonic mass and 2 pc for the
models with 10$^7$ M$_\odot$ of baryons.  The resolution decreases
outwards with a ratio between adjacent cells of 1.02.  The reference
values for some key parameters (common to all 18 basic models) are
summarized in Table \ref{table:1} whereas the specific parameter
values distinguishing those models are recalled in Table
\ref{table:2}.  In order to appreciate the distinction between models
with different geometries, Fig. \ref{fig:init} shows the initial gas
distribution of the models L8R (upper panel), L8M (central panel) and
L8F (lower panel).

\begin{table*}
\caption{Parameter values for individual models and some representative 
results.}
\label{table:2}      
\centering                          
\begin{tabular}{c c c c c c c c c c c c c c}       
\hline 1 & 2 & 3 & 4 & 5 & 6 & 7 & 8 & 9 & 10 & 11 & 12 & 13 & 14\\
\hline\hline
Model & $M_b$$^{(a)}$ & $b/a$$^{(b)}$ & $f_g$$^{(c)}$ & 
$f_N$$^{(d)}$ & $M_{DM}$$^{(e)}$ & $r_{vir}$$^{(f)}$ & $\rho_0$$^{(g)}$ 
& $\epsilon_{gas, 200}$$^{(h)}$ &$\epsilon_{O, 200}$$^{(i)}$     
& $\epsilon_{gas, 500}$$^{(j)}$ &$\epsilon_{O, 500}$$^{(k)}$ 
& $A(O)_{200}$$^{(l)}$
& $A(O)_{500}$$^{(m)}$
\\
\hline                        
L7F & 10$^7$ & 0.2 & 0.6 & 0.5 & 2.41 & 1.43 & 1.36  
& 0.000 & 0.980 & 0.999 & 0.993 & 9.84 & 9.84\\      
L7M & 10$^7$ & 1.0 & 0.6 & 0.5 & 2.41 & 1.43 & 0.85  
& 0.999 & 0.978 & 0.999 & 0.993 & 9.85 & 9.81\\      
L7R & 10$^7$ & 5.0 & 0.6 & 0.5 & 2.41 & 1.43 & 0.39  
& 0.993 & 0.945 & 0.999 & 0.991 & 8.74 & 9.64\\      
L8F & 10$^8$ & 0.2 & 0.6 & 0.5 & 12.4 & 2.47 & 2.26  
& 0.365 & 0.812 & 0.606 & 0.826 & 7.23 & 7.84\\      
L8M & 10$^8$ & 1.0 & 0.6 & 0.5 & 12.4 & 2.47 & 1.60  
& 0.296 & 0.111 & 0.598 & 0.872 & 7.99 & 7.73\\      
L8R & 10$^8$ & 5.0 & 0.6 & 0.5 & 12.4 & 2.47 & 0.90
& 0.192 & 0.001 & 0.558 & 0.599 & 7.99 & 8.21\\      
L9F & 10$^9$ & 0.2 & 0.6 & 0.5 & 63.5 & 4.26 & 11.0  
& 0.041 & 0.006 & 0.183 & 0.639 & 7.87 & 7.90\\      
L9M & 10$^9$ & 1.0 & 0.6 & 0.5 & 63.5 & 4.26 & 5.50  
& 0.031 & 0.001 & 0.116 & 0.329 & 7.88 & 8.14\\      
L9R & 10$^9$ & 5.0 & 0.6 & 0.5 & 63.5 & 4.26 & 3.45  
& 0.059 & 0.000 & 0.123 & 0.163 & 7.90 & 8.26\\      
H7F & 10$^7$ & 0.2 & 0.9 & 2.0 & 3.22 & 1.58 & 1.50  
& 0.883 & 0.938 & 0.586 & 0.965 & 7.11 & 6.71\\      
H7M & 10$^7$ & 1.0 & 0.9 & 2.0 & 3.22 & 1.58 & 0.83  
& 0.860 & 0.824 & 0.833 & 0.943 & 7.48 & 7.31\\      
H7R & 10$^7$ & 5.0 & 0.9 & 2.0 & 3.22 & 1.58 & 0.47  
& 0.412 & 0.007 & 0.627 & 0.703 & 7.60 & 7.68\\      
H8F & 10$^8$ & 0.2 & 0.9 & 2.0 & 16.5 & 2.72 & 2.45  
& 0.057 & 0.001 & 0.310 & 0.580 & 7.41 & 7.60\\      
H8M & 10$^8$ & 1.0 & 0.9 & 2.0 & 16.5 & 2.72 & 1.50  
& 0.085 & 0.000 & 0.212 & 0.375 & 7.42 & 7.69\\      
H8R & 10$^8$ & 5.0 & 0.9 & 2.0 & 16.5 & 2.72 & 0.94  
& 0.068 & 0.000 & 0.170 & 0.061 & 7.42 & 7.84\\      
H9F & 10$^9$ & 0.2 & 0.9 & 2.0 & 84.7 & 4.69 & 6.00  
& 0.025 & 0.000 & 0.058 & 0.086 & 7.40 & 7.77\\      
H9M & 10$^9$ & 1.0 & 0.9 & 2.0 & 84.7 & 4.69 & 3.80  
& 0.013 & 0.000 & 0.048 & 0.001 & 7.39 & 7.81\\      
H9R & 10$^9$ & 5.0 & 0.9 & 2.0 & 84.7 & 4.69 & 2.45  
& 0.023 & 0.000 & 0.042 & 0.000 & 7.40 & 7.81\\      
\hline                                   
\end{tabular}
\tablefoot{$^{(a)}$: Initial baryonic mass in M$_\odot$ (within 0.5 $\cdot$ 
  $r_{vir}$);
  $^{(b)}$: Ratio between the scale lengths $a$ and $b$ in the initial 
  Miyamoto-Nagai stellar distribution (see Eq. \ref{eq:mn});
  $^{(c)}$: Gas-to-baryon fraction; $^{(d)}$: Ratio between the mass of 
  newly formed stars at the end of the SF period and the mass 
  of the pre-existing disk $M_d$; $^{(e)}$: Mass of the DM halo (in 
  10$^8$ M$_\odot$); $^{(f)}$: Virial radius (in Kpc, see eq. \ref{eq:rvir}); 
  $^{(g)}$: Central gas 
  density (in 10$^{-24}$ g cm$^{-3}$); $^{(h)}$: Ejected gas fraction after 
  200 Myr; $^{(i)}$: Ejected oxygen fraction after 200 Myr; $^{(j)}$: 
  Ejected gas fraction after 500 Myr; $^{(k)}$: Ejected oxygen fraction 
  after 500 Myr; $^{(l)}$: Abundance of oxygen measured as 12+log(O/H), 
  where O/H is the abundance ratio in number, at 200 Myr;
  $^{(m)}$: Abundance of oxygen measured as 12+log(O/H), 
  where O/H is the abundance ratio in number, at 500 Myr.
} 
\end{table*}

\section{Results}
\label{sec:results}

\subsection{The reference models}
\label{subs:refm}

\begin{figure*}
\centering
\begin{tabular}{cc}
\includegraphics[width=9cm]{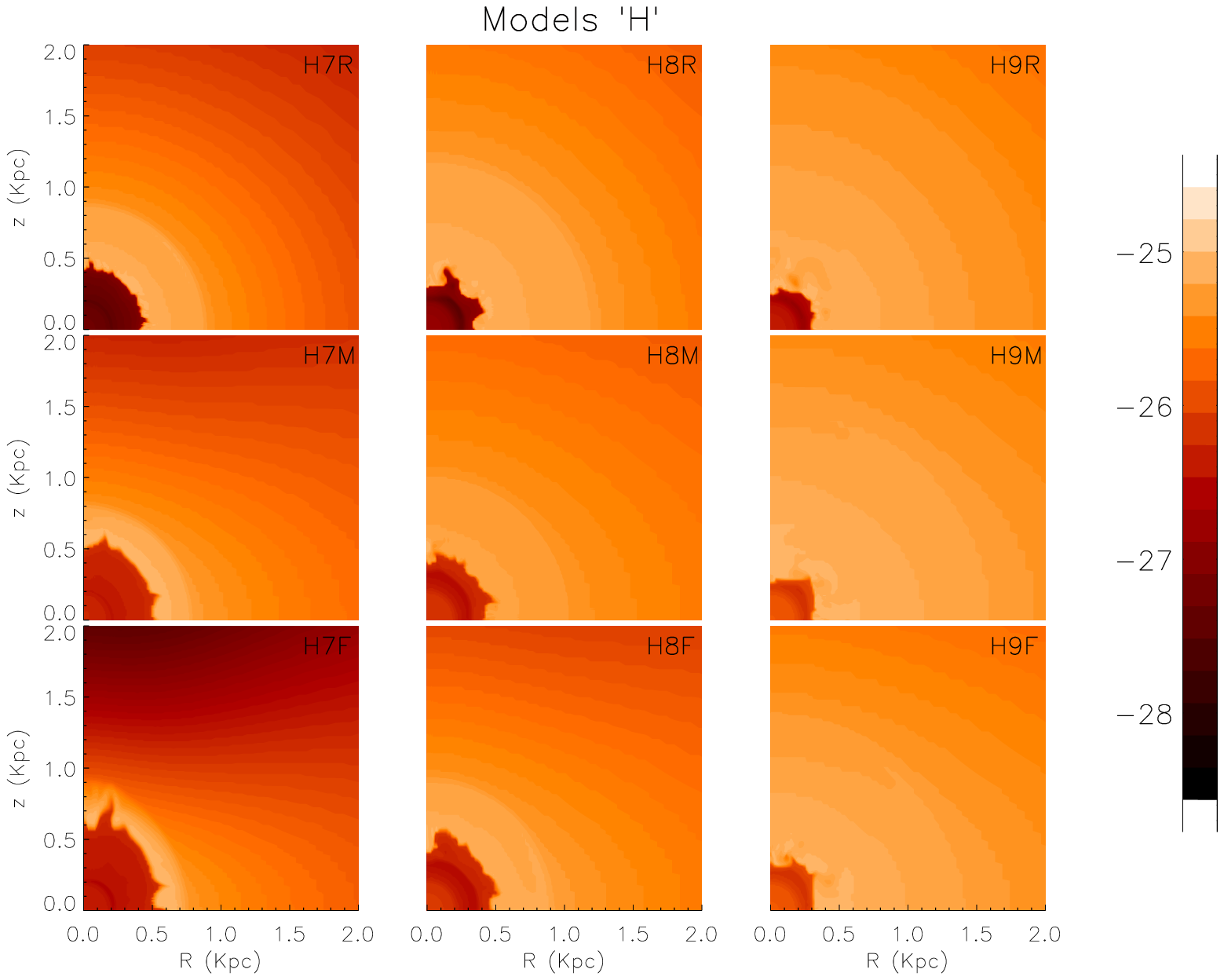} &
\includegraphics[width=9cm]{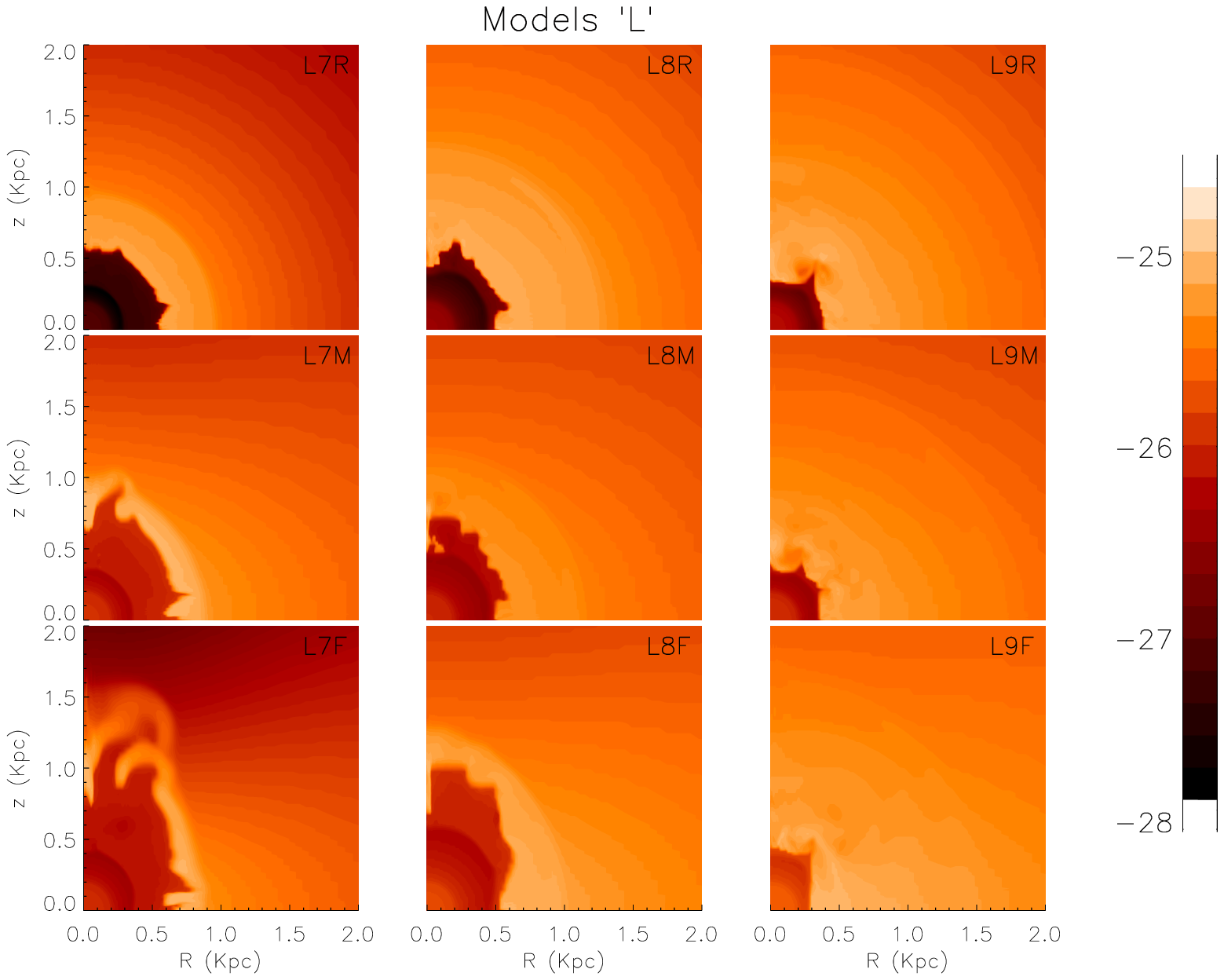}
\end{tabular}
\caption{Gas density distribution for the 18 reference models (see
  Table \ref{table:2}) after 100 Myr of evolution.  The 9 panels on
  the left represent the models with high initial gas fraction (models
  ``H''); the 9 on the right are the corresponding models with low
  initial fraction of gas (models ``L'').  For each sequence of 9
  panels, the first column represents models with 10$^7$ M$_\odot$ of
  initial baryonic mass; the middle column shows the gas distribution
  for models with mass 10$^8$ M$_\odot$ and finally the on the right
  column the models with 10$^9$ M$_\odot$ are displayed.  The top rows
  of models are characterized by a roundish initial distribution
  (models ``R''; with $b/a=5$).  The middle rows show models with
  $b/a=1$ (models ``M'') and finally the bottom rows are characterized
  by $b/a=0.2$ (flat models or ``F'').  At the top-right corner of
  each panel the model designation is also indicated.  For each set of
  9 models, the left-hand strip shows the (logarithmic) density scale
  (in g cm$^{-3}$).}
\label{fig:snap100}
\end{figure*}

\begin{figure*}
\centering
\begin{tabular}{cc}
\includegraphics[width=9cm]{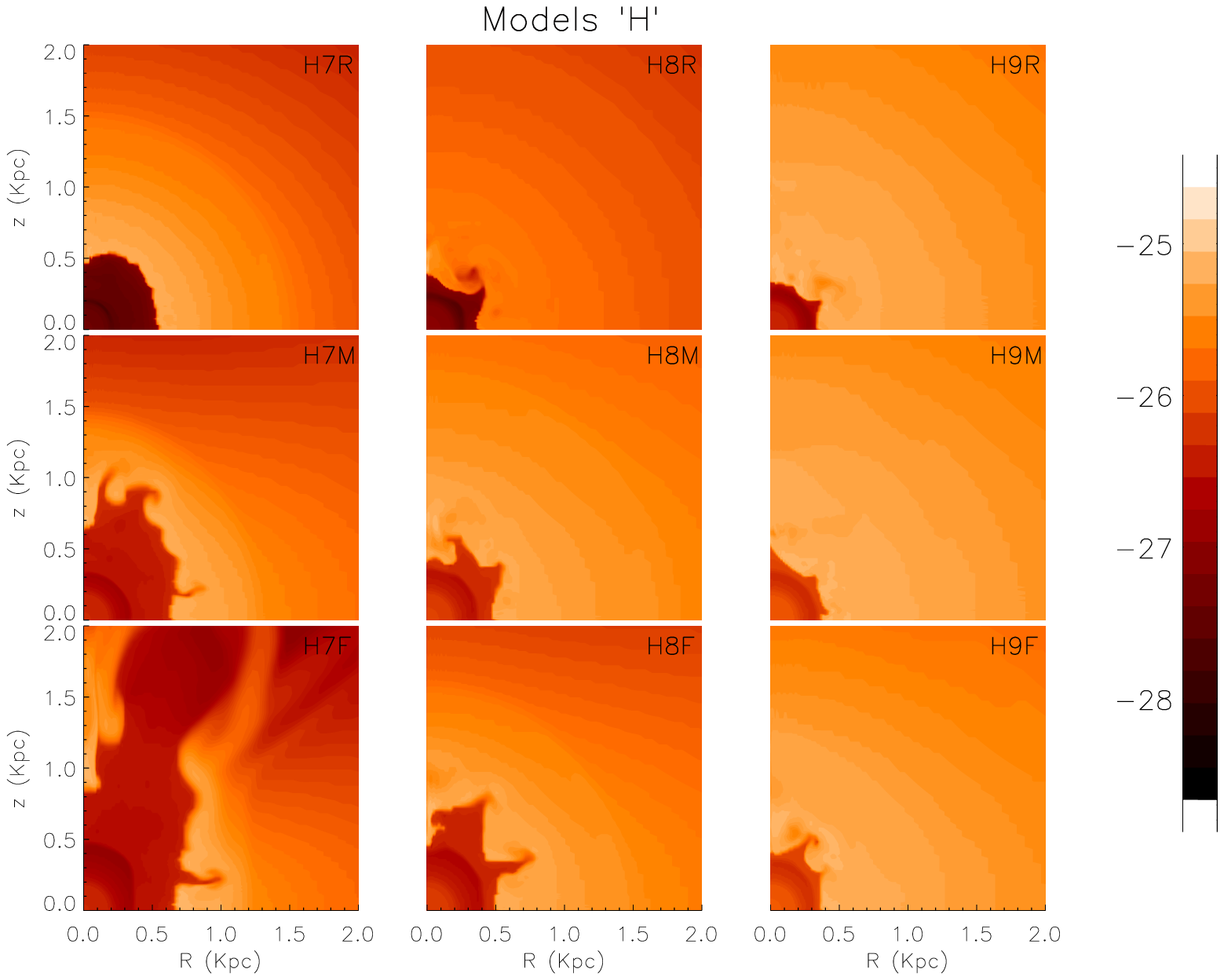} &
\includegraphics[width=9cm]{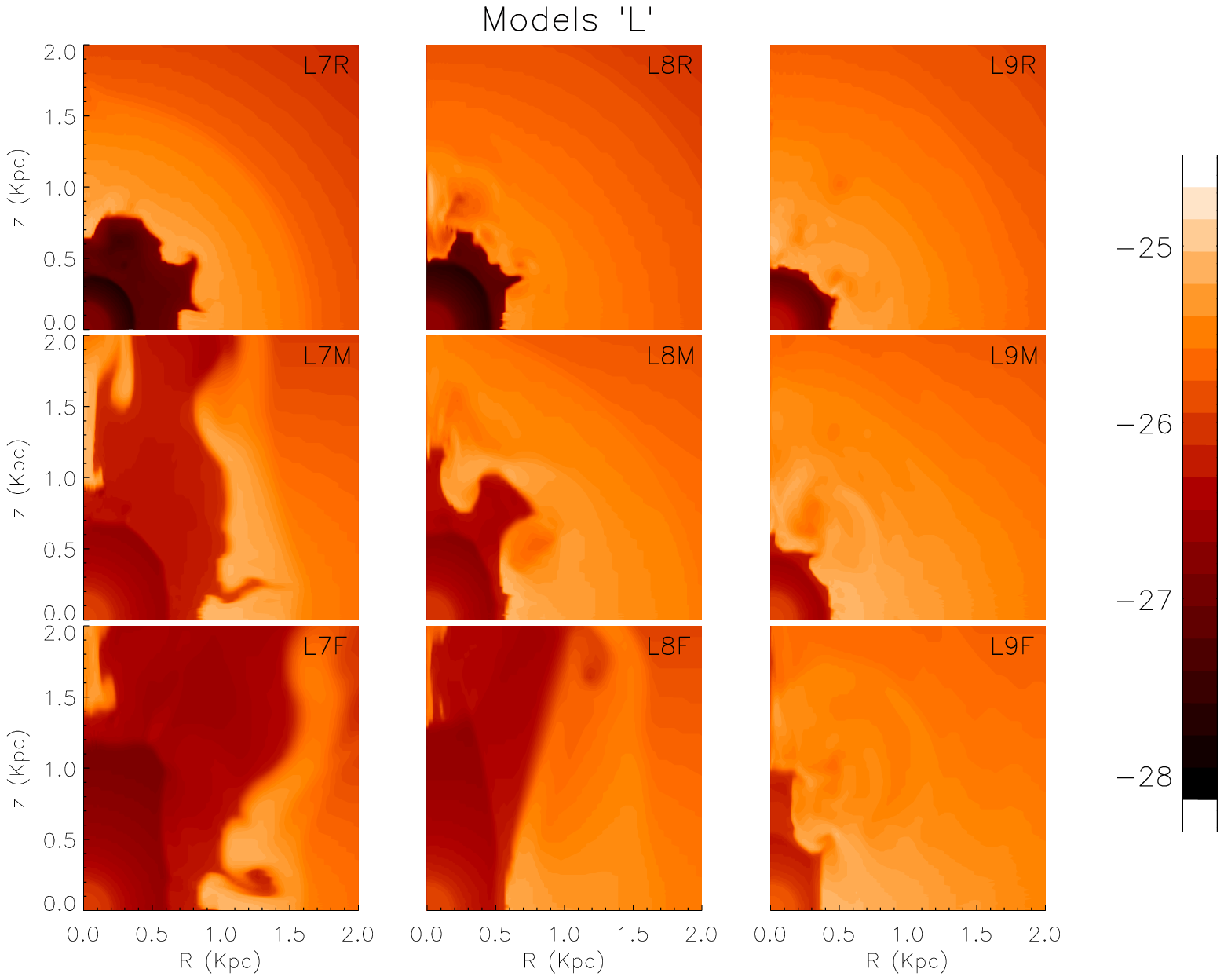}
\end{tabular}
\caption{Same as Fig. \ref{fig:snap100} but for snapshots taken after 
  200 Myr of galactic evolution.}
\label{fig:snap200}
\end{figure*}

\begin{figure}
\centering
\includegraphics[width=9cm]{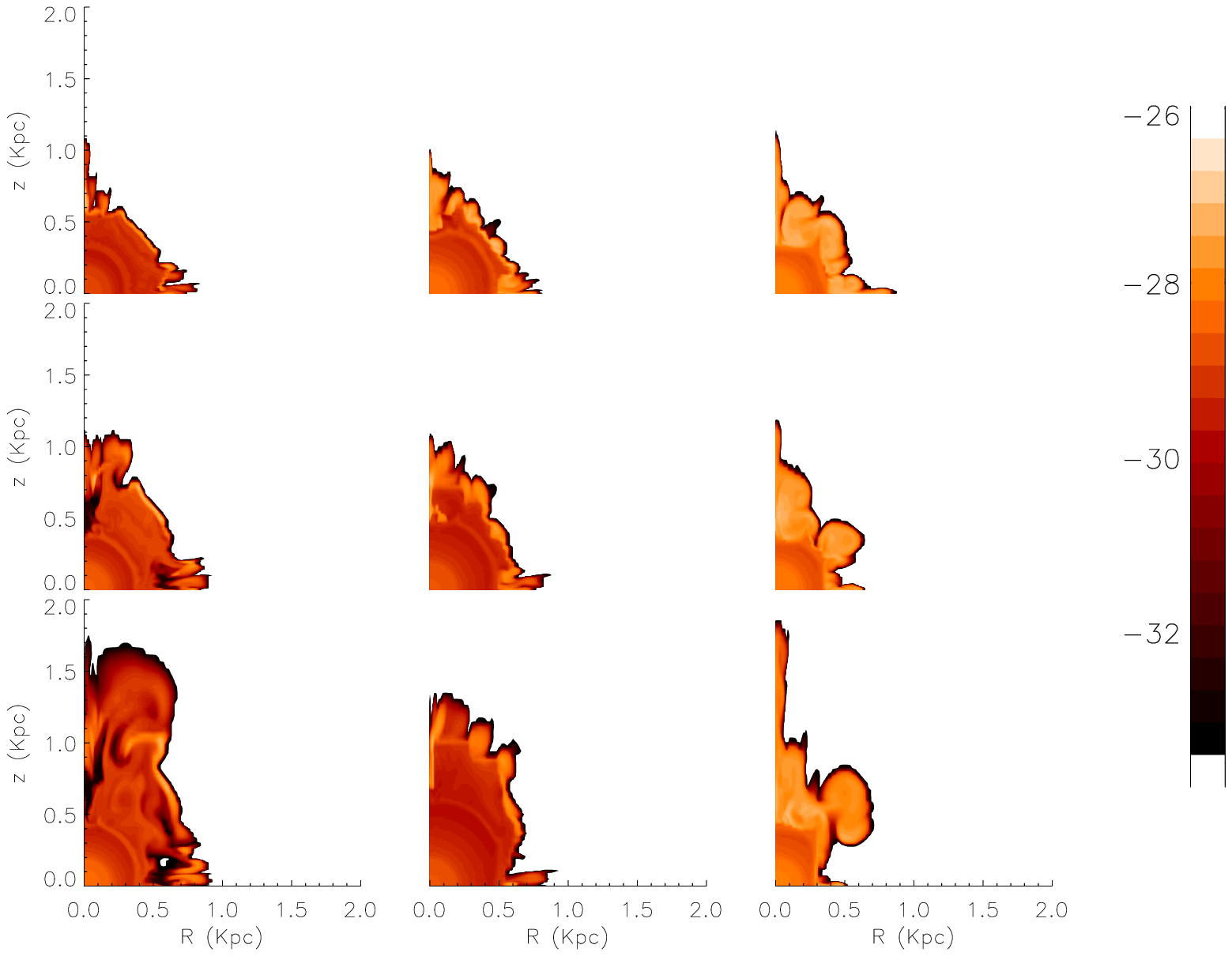}
\caption{Distribution of oxygen mass density after 100 Myr of
  evolution for the 9 models characterized by a small initial gas
  fraction (models ``L'' in Table \ref{table:2}).  The oxygen density
  scale (in g cm$^{-3}$) is on the left-hand strip}
\label{fig:ox100}
\end{figure}

A snapshot of the evolution of all 18 reference models (see Table
\ref{table:2}) is shown in Fig. \ref{fig:snap100} (after an
evolutionary time of 100 Myr) and in Fig. \ref{fig:snap200} (after an
evolutionary time of 200 Myr).  The main features of our model results 
can be noticed from these two figures:
\begin{itemize}
\item As expected, models characterized by a flat initial gas
  distribution (bottom rows) more easily develop a large-scale
  outflow.  The roundish models (top rows) show very weak signs of
  outflows even at 200 Myr, irrespective of the baryonic mass.
\item For models characterized by the same geometry, the initial
  baryonic mass clearly plays a very important role, with much more 
  prominent outflows in low-mass galaxies (left-most columns for each 
  group of panels).  This confirms the overall trend of gas ejection 
  efficiency vs. mass shown by MF99 for instance.
\item If a large-scale outflow is formed, freshly produced metals can
  be easily lost from the galaxy (it is to remind that at an
  evolutionary time of 100 or 200 Myr the SF is still ongoing).  This
  is better seen in Fig. \ref{fig:ox100} where the distribution of
  oxygen (taken as a proxy of metals) is shown for the models with
  small initial gas fractions (models ``L''), after 100 Myr.  It is
  clear from this figure that in many model galaxies a large fraction
  of oxygen is leaving the parent galaxy.  This will be quantified in
  detail below.
\item Outflows in models with high initial gas fraction (models ``H'';
  the ones on the left of Figs. \ref{fig:snap100} and
  \ref{fig:snap200}) are clearly less prominent compared to the
  corresponding outflows in low-gas models.  This is due to the fact
  that the expansion of superbubbles is hampered by larger gas
  pressures in these models and to the fact that the SFRs are lower.
\end{itemize}

\begin{figure}
\centering
\includegraphics[angle=-90,width=9cm]{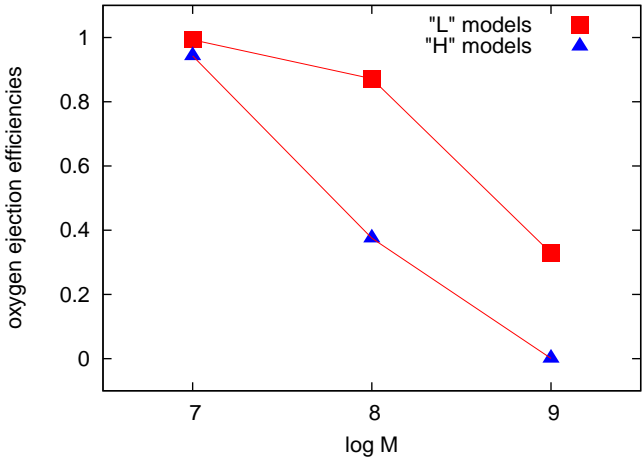}
\caption{Oxygen ejection efficiencies after 500 Myr of evolution for
  medium ($b/a=1$) models as a function of the initial baryonic mass.
  Squares refer to models with initially low gas fractions and
  triangles to models with high $f_g$.}
\label{fig:eff1}
\end{figure}

\begin{figure}
\centering
\includegraphics[angle=-90,width=9cm]{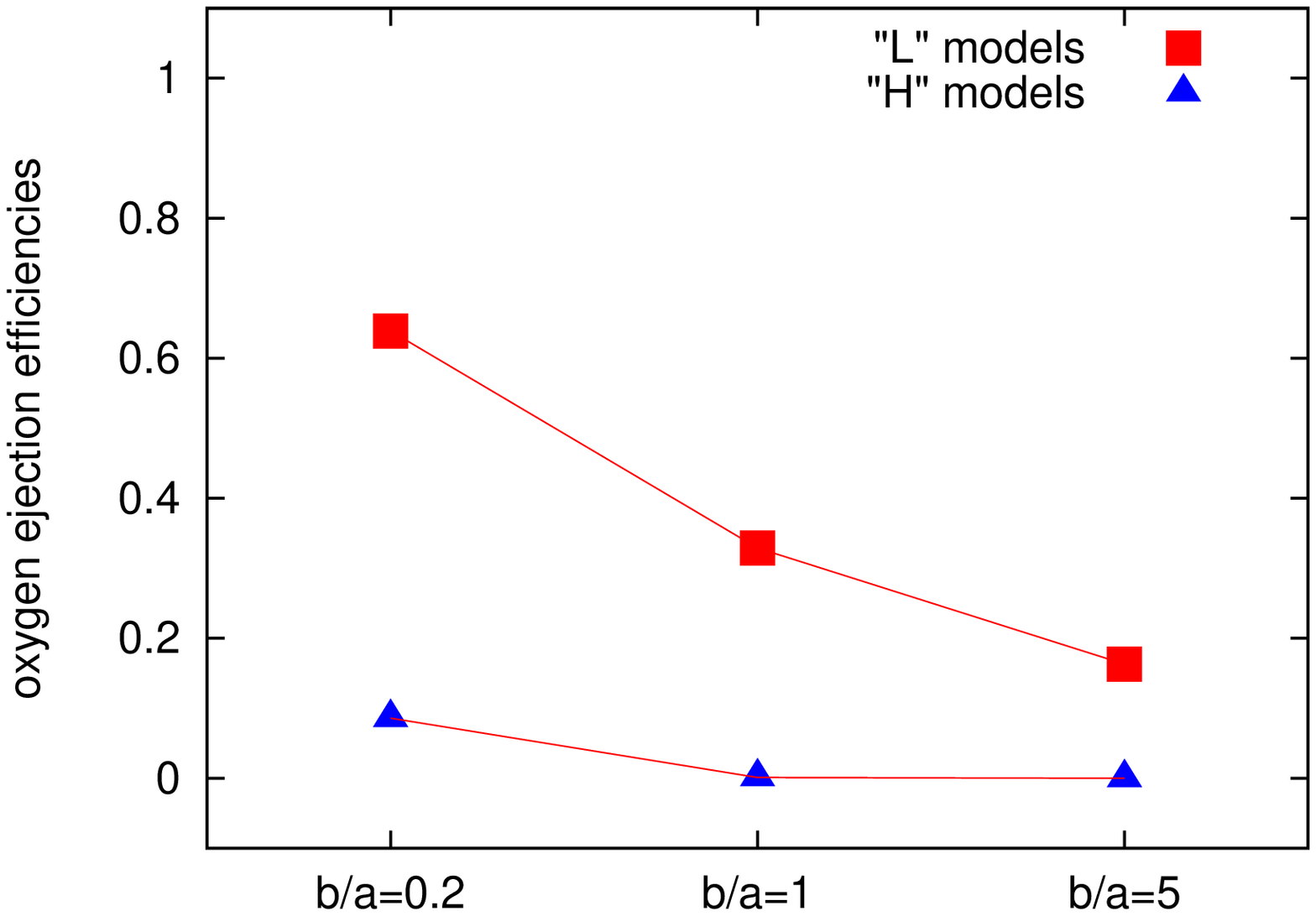}
\caption{Oxygen ejection efficiencies after 500 Myr of evolution for
  high-mass ($M_b=10^9$ M$_\odot$) models as a function of the degree
  of flattening $b/a$.  Symbols as in Fig. \ref{fig:eff1}.}
\label{fig:eff2}
\end{figure}

A more quantitative analysis of the results of the models is obtained
by inspecting the columns 9--12 of Table \ref{table:2}.  Here the
ejected fractions $\epsilon$ of gas and oxygen are indicated for each
model after 200 and 500 Myr.  We simply estimate the retained gas
fractions as the ratios between the mass of pristine gas contained in
0.5 $r_{vir}$ at the time 200 (500) Myr and the same mass at the
beginning of the simulation.  To calculate the retained oxygen
fraction we divide instead the mass of oxygen within 0.5 $r_{vir}$ by
the total amount of oxygen expelled by dying stars until 200 (500)
Myr.  The ejection efficiencies (of gas and oxygen) can thus be simply
estimated as $1-rf$, where $rf$ indicates the corresponding retained
fraction.  Although our estimate is approximate and can be affected
for instance by gas (or oxygen) temporarily leaving the galaxy and
being re-accreted later on, it is clear that estimates based on the
escape velocities (as in MF99) are approximate as well (ejected gas
does not evolve ballistically), thus we keep our simple definition of
ejected fractions, bearing in mind the inherent uncertainties, thus
focusing on differences in the ejected fractions rather than on
absolute values.  A close inspection of the ejected fractions reveal
the following properties of the models:
\begin{itemize}
\item In roundish models the ejected fractions of oxygen tend to be
  close or slightly smaller than the ejected fractions of pristine
  gas.  This is due to the fact that these models do not largely
  depart from their initial spherical symmetries.  Under these
  circumstances, most of the metals remain confined inside the (almost
  spherical) superbubbles (the darker central regions in the upper
  rows of Figs. \ref{fig:snap100} and \ref{fig:snap200}; see also Fig.
  \ref{fig:ox100}).  A large fraction of the pristine gas is swept-up
  in a relatively dense shell surrounding the superbubble (the
  supershell) and, occasionally, a fraction of this gas can be located
  at distances from the galactic center greater than 0.5 $\cdot$
  $r_{vir}$.
\item On the contrary, non-spherical models (models ``M'' and ``F'')
  clearly show a tendency of retaining pristine gas more easily than
  metals.  This is in agreement with many previous studies already
  mentioned in the Introduction and in Sec. \ref{sec:summ} (e.g. MF99;
  D'Ercole \& Brighenti \cite{db99}; Recchi et al. \cite{rmd01}).
\item If we focus on models with the same initial gas distribution, we
  recover a clear trend of decreasing metal ejection efficiencies with
  increasing galactic baryonic mass, once again in agreement with the
  results of MF99.  For instance, the oxygen ejection efficiencies at
  the end of the simulations (i.e. $\epsilon_{O, 500}$) for the ``M''
  models are shown in Fig. \ref{fig:eff1} (squares: models with low
  initial gas fractions; triangles: models with initial high gas
  fractions).  Oxygen ejection efficiencies are close to 1 for models
  with 10$^7$ M$_\odot$ of baryonic mass and very low for models with
  $M_b=10^9$ M$_\odot$.
\item As already noticed, models with a larger initial gas fraction
  are less effective in developing galactic winds.  Consequently, the
  ejected fractions for these models are systematically lower than
  the corresponding ``L'' models (the ones with a lower initial gas
  fraction -- see also Fig. \ref{fig:eff1}).
\item The effect of geometry on the development of galactic winds and
  on the fate of metals (and also on the fate of pristine gas) is
  quite evident from this table: models with the same baryonic mass
  can show very different retained fractions depending on their
  degrees of flattening.  Model L8R for instance retains almost all
  the oxygen at 200 Myr ($\epsilon_{O,200}=0.001$).  The model L8F,
  with the same baryonic mass but with a flat initial distribution
  retains only 18.8\% of oxygen, i.e. it expels 81.2\% of the produced
  oxygen through galactic winds.  If we look at the final ejection
  efficiencies, we can notice again that the roundish model has the
  smallest oxygen ejection efficiency.  The models L8F and L8M have
  both quite low retained oxygen fractions (therefore high oxygen
  ejection efficiencies), although $\epsilon_{O, 500}$ of model L8F is
  slightly smaller than the ejected fraction of L8M.  The final
  retained gas fractions for all models depend very little on the
  degree of flattening.  In contrast, focusing on the ``H8'' family of
  models (high gas fraction, 10$^8$ M$_\odot$ of initial baryonic
  mass), we can notice again that the final ejection efficiencies of
  gas depend very little on $b/a$, whereas the final oxygen ejection
  efficiencies are 0.58, 0.375 and 0.061 for the models H8F, H8M and
  H8R, respectively (see column 12 of Table \ref{table:2}), i.e. the
  metal ejection efficiencies can change by up to one order of
  magnitude depending on the geometry.  Similar trends with the degree
  of flattening (little effect on the gas ejection efficiencies, large
  effect on the metal ejection efficiencies) can be noticed for the
  families of models ``L9'', ``H9'' and ``H7'', too.  The comparison
  of the final oxygen ejection efficiencies for the families of models
  ``L9'' and ``H9'' is shown in Fig. \ref{fig:eff2}
\end{itemize}

The last two columns of Table \ref{table:2} show the oxygen abundance
measured as 12+log(O/H)abundance ratio, where O/H is the abundance
ratio in number, of all model galaxies.  This abundance ratio is a
mass-weighted average of the abundance ratios in each computational
cell lying within a sphere of radius 0.5 $\cdot$ $r_{vir}$.  Again,
this is only an approximate measure of the galactic metallicity and
only relative abundance ratios between different model galaxies are
relevant. In particular, the present grid of models is not aimed at
reproducing the observed metallicities of individual DGs and it can be
noticed e.g., that large galaxies do not always show larger O/H ratios
compared to smaller galaxies (at variance with the observed M-Z
relation; see Lee et al. \cite{lee06}).  It is also worth noticing
that, unexpectedly, the models with the highest metallicities are the
ones experiencing the strongest galactic winds (models ``L7''), at
variance with the general idea that galactic winds can keep the
metallicities of galaxies low (see the Introduction).  This is due to
the fact that these model galaxies expel almost all pristine gas (see
Table \ref{table:2}) and therefore the (continuously produced) oxygen
is mixed with a very small fraction of unpolluted gas.  The average
metallicity of the galaxy is thus close to the (very large)
metallicity of the stellar ejecta.  However, this gas is deemed at
leaving the parent galaxy and is anyway too hot to form star; it will
thus not increase the average metallicity of the stellar populations
in the galaxy.  The models attaining the lowest metallicities are the
ones (like H7F) developing large-scale outflows but retaining a large
fraction of pristine gas bound to the galaxy.  See also Sect.
\ref{sec:disc} for a more extended discussion on the connection
between the results of our models and the M-Z relation.

\subsection{Widening the parameter space}
\label{subs:parspace}

As already mentioned in Sect. \ref{subs:setup}, we have considered
also models with different SFHs, different feedback radii, different
initial metallicities and with perturbed initial distributions of gas.
We summarize in this subsections how these parameters affect our
findings.

\subsubsection{Dependence on the star formation history}
\label{subs:SFH}
We have re-run some models (specifically the ones for the family
``L8'') in which the SFH, instead of being constant for 500 Myr as in
the standard models, is different from zero only during the first 50
Myr, but the SFR is ten times more intense than in the standard
models.  A large-scale outflow quickly develops, even for the
roundish model, sustained by the large rate of SNII explosions during
the first $\sim$ 80 Myr.  At the end of this phase, energy is still
supplied by Type Ia SNe and winds from intermediate-mass stars, but at
a lower rate.  This energy is still able to sustain the outflow for
some time, but after a few times 10$^8$ yr, the gravitational pull
prevails over the pressure gradient caused by the hot cavity of gas
and some gas begins to recollapse towards the center of the galaxy
(see Recchi \& Hensler \cite{rh06} for a precise assessment of this
phenomenon).  

The result is that the calculated ejection efficiencies (of gas and
metals) are larger at 100 or 200 Myr than at 500 Myr.  In particular,
the ejection efficiencies at 200 Myr are much larger than the ones in
the standard models, tabulated in Table \ref{table:2}.  Disk-like
models (models ``F'' and ``M'') retain only a few percent of gas and
metals, and also the roundish model (``R'') is able to expel more than
50\% of the initial gas at distances larger than 0.5 $\cdot$
$r_{vir}$.  The large gas ejection efficiencies for these models are
due to the fact that, at these high rates of energy release, also the
horizontal transport of gas can be effective and some gas is pushed to
large distances from the galactic center even along the disk.  The
disk-like ``M'' and ``F'' models show again metal ejection
efficiencies (slightly) larger than gas ejection efficiencies.
Consequently, the metallicities of these model galaxies are quite low
during the SF episode.

However, as soon as the luminosity considerably reduces, this gas is
pulled again within 0.5 $\cdot$ $r_{vir}$ (i.e. a fall back is
occurring), whereas the material previously channelled along the
galactic wind accelerates away of the galaxy due to the steep density
gradient.  Consequently, after 500 Myr of evolution, the retained gas
fractions for these models tend to be larger (by a factor of $\sim$
1.5--2) than the corresponding fractions of the standard models (the
ones reported in column 11 of Table \ref{table:2}), whereas the
retained fraction of metals are only slightly larger.

\subsubsection{Dependence on the feedback radius}
We have considered (again for the family of models ``L8'') a variation
of the radius over which energy and metals are injected.  As already
anticipated in Sect. \ref{subs:setup}, instead of the reference value
$R_F=$ 200 pc, we have calculated models with $R_F=50$ pc and
$R_F=1000$ pc.  The models with $R_F=1000$ show very large ejection
efficiencies for both gas and metals.  This is due to the unfortunate
circumstance that 0.5 $\cdot$ $r_{vir}$ for this family of models is
only slightly larger than 1000 pc, therefore a moderate energy
injection already suffices to push a large fraction of gas at
distances larger than 0.5 $\cdot$ $r_{vir}$.  The comparison with the
ejection efficiencies of standard models makes thus little sense.  We
have therefore calculated the ejection efficiencies of these models
(and of the corresponding standard models) at $r=r_{vir}$.  These turn
out to be $\sim$ 10\% smaller than the ones of corresponding reference
models.  We confirm therefore that, if the energy is redistributed
over a large volume, radiative energy losses are more effective and
the final ejection efficiencies are reduced, in agreement with the
findings of Fragile et al.  (\cite{frag04}).  Also the models with
$R_F=50$ pc show slightly lower ejection efficiencies.  In this case,
the enhanced cooling efficiency is due to the fact that the gas within
the feedback radius is now characterized by large densities and large
metallicities (see also Tenorio-Tagle et al. \cite{tt07}).  Also in
this case, the effect of the feedback radius is quite limited
(ejection efficiencies change by $\sim$ 10--15 \%).  As also shown by
Rodr{\'{\i}}guez-Gonz{\'a}lez et al. (\cite{rodr11}; see their fig. 7)
the dependence of the ejection efficiencies on the feedback radius is
non-monotonic, although the dependence of our results on this
particular parameter is quite limited.

\subsubsection{Dependence on the initial metallicity}
Since we consider models in which a stellar disk is already present at
the beginning of the simulations, it seems unreasonable to start with
a primordial metallicity.  We considered therefore models in which the
initial metallicity (of stars and gas) is regulated by the mass of the
pre-existing disk, namely, according to the M-Z relation, the larger
the disk mass, the larger the initial metallicity we must consider.
In particular, if we take into account the correlation between mass
and metallicity obtained by Tremonti et al. (\cite{trem04}), models of
the ``L8'' family should have an initial metallicity of about one
tenth of solar.  Of course, the final metallicities of these models
will be much higher than the ones attained by models initially without
metals.  However, as already mentioned, the only dynamical effect of a
different initial metallicity is to increase the radiative losses and
thus to reduce the amount of thermal energy of the galaxy.  It turns
out that this reduction is quite limited and, therefore, the ejection
efficiencies do not change substantially (they are only a few per cent
lower than the ejection efficiencies of corresponding models initially
without metals).  For these models, the metal ejection efficiencies
are calculated according only to the amount of freshly produced metals
retained or ejected.

\subsubsection{Models with random perturbation of the initial gas
  distribution}
\label{subs:random}
\begin{figure}
\centering
\vspace{-2cm}
\includegraphics[width=9.5cm]{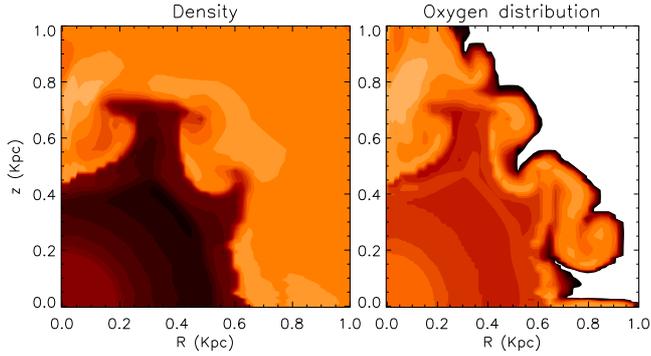}
\caption{Distribution of gas (left panel) and oxygen (right panel)
  density for a model with large perturbation of the initial density
  distribution (see text), after 200 Myr of evolution.  Brighter
  colors correspond to larger densities.}
\label{fig:dox}
\end{figure}

As described in Sect. \ref{subs:setup}, once an equilibrium initial
configuration has been obtained, we have perturbed it by means of eq.
\ref{eq:pert}.  In particular, we have perturbed the model L8R by
amplitudes $\varepsilon$ of 1\% (mildly perturbed model) and 5\%
(largely perturbed model).  The perturbation increases the turbulence
of the model.  Mixing between hot and cold regions is enhanced and the
net effect is an increase in the radiative energy losses, hence a
reduced amount of energy available to drive galactic winds (in other
words, the galactic winds become more mass loaded).  As a consequence
of that, the retained gas fraction of the mildly perturbed model
increases (from the final value of 0.442 of the model L8R it increases
to 0.547).  However, the final retained fraction of freshly produced
metals (slightly) decreases for this model (it reduces to 0.345 from
the reference value of 0.401).  This is due to the fact that it is
much more difficult to keep the model close to spherical symmetry if
the initial gas distribution is perturbed.  Regions of lower pressure
can be created randomly within the computational box and the
propagation of metals can thus deviate from isotropicity, being faster
(and allowing some venting out of metals) along (randomly oriented)
directions with steep pressure gradients.  This effect is similar to
the one analyzed in great detail by RH07 (see also Sect.
\ref{sec:summ} for a short summary of the main results of this paper).
In the largely perturbed model, the final metal ejection efficiency is
similar to the one obtained in the mildly perturbed model.  However,
the venting out (along random directions) involves now also a
non-negligible fraction of pristine gas and the final gas retained
fraction is 0.443 (very similar to the gas retained fraction of the
reference model L8R).  The piercing of the supershell along different
directions and not just along the direction perpendicular to the disk
of the galaxy can be appreciated by inspecting Fig. \ref{fig:dox}.  It
shows the gas and oxygen density distribution after 200 Myr of
evolution of the largely perturbed model.  Clearly, the perturbation
of the initial gaseous distribution affects the ejection efficiencies
of gas and metals in a non-linear (and some times non-predictable) way
and deserves further studies.

It is important to stress that, due to the assumed symmetry and
dimensionality of the code, inhomogeneities represent rings of denser
gas.  It is not clear whether a realistic 3-D distribution of
inhomogeneities will produce the same results.  Our group is currently
running (with the AMR code FLASH) 3-D chemo-dynamical simulations of
DGs with inhomogeneities (Mitchell et al., in preparation).  The
results of these simulations can shed light on the reliability of our
results.

\section{Discussion and conclusions}
\label{sec:disc}
In this paper we have studied the development of galactic outflows in
DG models.  Our main focus is to study the dependence of the ejected
mass fractions (of gas and freshly produced metals) on the degree of
flattening of the galaxy, a problem that has received little attention
in the past.  A very simple theoretical analysis leads to the
conclusion that, in a very flat galaxy, the ejection efficiencies of
freshly produced metals must be quite high.  In fact, once the
over-pressurized gas formed by SN explosions and stellar wind breaks
out and forms a bipolar galactic wind, metals can be easily ejected
out of the galaxy through this funnel.  On the other hand, in a
spherical (or almost spherical) galaxy, bipolar outflows cannot be
formed because there is no preferential direction, along which the
pressure gradient is steeper.  Therefore, either the galaxy expels gas
isotropically (but blow-away is very difficult in DGs, see e.g.
Hensler et al.  \cite{hens04}) or the superbubble of hot (and
metal-enriched) gas remains confined inside the galaxy (thus the metal
ejection efficiencies are very low).  We have considered three
different degrees of flattening for our model galaxies: a roundish
model (one for which the ratio between the length scales $a$ and $b$
describing the Miyamoto-Nagai potential is $b/a=5$), a thick-disk
model (with $b/a=1$) and a thin disk model (with $b/a=0.2$).  Our
study confirms the trend described above: ejection efficiencies in
roundish galaxies are systematically lower than the ones in disk-like
galaxies (see Table \ref{table:2} or Fig. \ref{fig:eff2}).  On the
other hand, for most of the analyzed models, transport of gas along
the disk is quite limited, therefore (in agreement with many previous
studies, e.g.  MF99; D'Ercole \& Brighenti \cite{db99}) we can
conclude that ejection efficiencies of gas are (at least for disk-like
models) lower than ejection efficiencies of freshly produced metals.

\begin{figure}
\centering
\includegraphics[angle=-90,width=9cm]{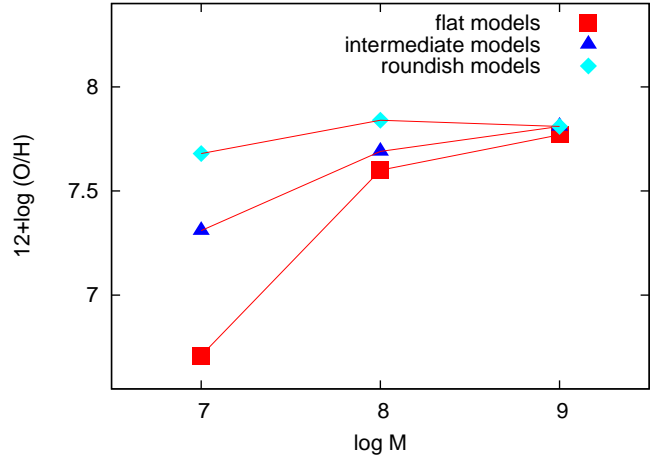}
\caption{Oxygen abundances (in number) after 500 Myr of evolution for
  high gas fraction models (models ``H'', with $f_g=0.9$), as a
  function of initial baryonic mass.  Squares refer to flat models
  (models with $b/a=0.2$); triangles show the results for intermediate
  models ($b/a=1$) and rhombs correspond to roundish models
  ($b/a=5$).}
\label{fig:oh}
\end{figure}

We have also studied in detail the effect of galactic mass on the fate
of gas and freshly produced metals, a problem that, on the contrary,
has received much more attention in the past (MF99;
Rodr{\'{\i}}guez-Gonz{\'a}lez et al. \cite{rodr11}).  We confirm that
smaller DGs, with shallower potential wells, favor the development of
large-scale outflows, hence the ejection efficiencies (of gas and
metals) increase with decreasing galactic masses (see Table
\ref{table:2} or Fig. \ref{fig:eff1}).  The increase of metal ejection
efficiencies through galactic winds has been often invoked to explain
the observed M-Z relation in galaxies (see e.g. Tremonti et al.
\cite{trem04}).  The dependence of the obtained metallicity (measured
by means of 12+$\log$ (O/H)) of our model galaxies as a function of
their initial masses is shown in Fig. \ref{fig:oh} for the ``H'' set
of models.  For all three degrees of flattening, a trend of increasing
metallicity as a function of mass is clearly visible.  However, as
outlined also in Sec. \ref{subs:refm}, for the ``L'' set of models
this trend is much less visible.

\begin{figure}
\centering
\includegraphics[angle=-90,width=9cm]{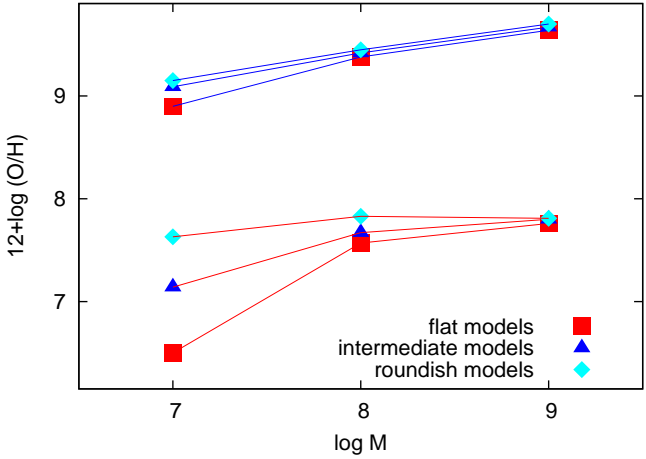}
\caption{Oxygen abundances (in number) after 500 Myr of evolution for
  ``H'', as a function of initial baryonic mass.  Symbols are as in
  Fig. \ref{fig:oh}.  The abundances in the cold (T $<$ 2$\cdot$10$^4$
  K) phase are shown in the lower set of points (connected by red
  lines).  The abundances in the hot (T $>$ 2$\cdot$10$^4$ K) phase
  are shown in the upper set of points (connected by blue lines).}
\label{fig:oh2}
\end{figure}

It is important to notice that the abundances tabulated in Table
\ref{table:2} and plotted in Fig. \ref{fig:oh} are mass-averaged means
within the whole galactic region.  However, two gas phases coexist
within this region: a hot gas phase (the galactic wind and the
starburst region) and a colder ISM.  We expect the hot phase to have
much higher metallicities than the cold phase.  The metallicity of the
galactic wind is expected to play no role in the process of chemical
enrichment of the galaxy (this gas is destined to leave the parent
galaxy) but is important in the process of chemical enrichment of the
intra-cluster medium.  To have an idea on the oxygen abundances of
these two phases, we distinguish grid points with temperatures above
and below 2 $\cdot$ 10$^4$ K.  Gas above this temperature threshold is
supposed to be too hot to be detected by optical spectroscopy.  The
resulting values of 12+log(O/H) after 500 Myr of evolution for the
``H'' family of models is shown in Fig.  \ref{fig:oh2} (red lines:
cold phase; blue lines: hot phase).  As expected, the abundances in
the cold phase are much lower than the abundances in the hot phase.
The cold phase abundances show the same trend of the total abundances
shown in Fig.  \ref{fig:oh}.  This is mainly due to the fact that the
calculated mean abundances are mass-weighted and most of the gas mass
is colder than 2 $\cdot$ 10$^4$ K.  For some models (in particular the
models with log M=7), the cold phase abundances are significantly
lower ($\sim$ 0.2 dex) than the total abundances.  This is expected
because, in some models, the metal-rich hot gas (wind and starburst
region) represents a significant fraction of the total gas.  On the
other hand, models experiencing no (or very limited) galactic winds
contain much less hot gas, therefore the average total abundance is
less affected by the metallicity of the hot phase and is much closer
to the abundance of the cold gas (the differences are a few hundredths
of dex).

\begin{figure}
\centering
\includegraphics[angle=-90,width=9cm]{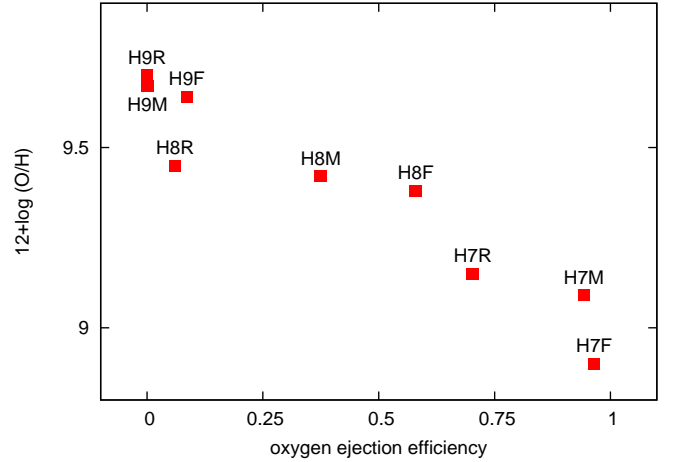}
\caption{Oxygen abundances in the hot phase plotted versus the oxygen
  ejection efficiencies (12th column in Table \ref{table:2}).  Model
  designations are labelled close to each point.}
\label{fig:ej_hotoh}
\end{figure}

Also the abundances of the hot phase tend to grow with mass (see the
blue lines in Fig. \ref{fig:oh2}).  High mass galaxies show higher hot
phase abundances because, for these models, much of the hot gas is in
the starburst region and thus its metallicity is close to the (very
high) metallicity of the stellar ejecta.  On the other hand, galaxy
models experiencing strong galactic winds show smaller hot-phase
metallicities because the winds entrain more cold (and metal-poor)
gas.  This process is usually indicated as mass-loading and is
confirmed by many studies (both theoretical and observational) of
galactic winds (see e.g. Strickland \& Stevens \cite{ss00}; Tescari et
al. \cite{teschio09}; Hopkins et al.  \cite{hopk12}; Newman et al.
\cite{newm12} among many others).  Indeed, a clearer correlation can
be shown between the ejected oxygen fraction after 500 Myr (12th
column in Table \ref{table:2}) and the hot-phase oxygen abundance.
This correlation is shown in Fig. \ref{fig:ej_hotoh}.  Models
experiencing galactic winds (thus with high oxygen ejection
efficiencies) show also mass-loading, thus dilution of the oxygen
abundances in the hot phase.  On the other hand, in models with weak
(or without) galactic winds most of the hot gas is concentrated in the
starburst region, where the abundances closely resemble the (very
high) abundances of the stellar ejecta.

One should warn the reader that the processes of metal dispersal and
mixing are very difficult to model.  No hydrodynamical code is able to
correctly resolve contact discontinuities that form between hot
cavities and cold shells.  Numerical diffusion tends to smear this
discontinuity and create regions of intermediate temperature and
densities, where metals can cool in a relatively short time-scale and
then mix with the surrounding shell.  In spite of these uncertainties,
we believe that 500 Myr is a sufficient time span to cool down most of
the produced metals (if they are not channelled in a galactic wind).
This is confirmed by some of our previous studies (see in particular
Recchi et al. \cite{rmd01}; their sect. 3.2) but also by other studies
addressing the specific issue of metal mixing (e.g. de Avillez \&
MacLow \cite{dm02}; Pan et al. \cite{pan12}; Yang \& Krumholz
\cite{yk12}), which confirm that a few hundred Myrs is a sufficient
time to disperse and mix metallicity inhomogeneities.  It is also
important to stress that, in our results, cooling and mixing of
stellar ejecta is mainly due to thermal conduction and to the
formation of eddies and vortices.  The presence of inhomogeneities
(see Sect. \ref{subs:random}) can facilitate both processes, thus
enhance the mass loading of the galactic winds.

One should also notice that the chosen temperature threshold
separating the hot and the cold phase (2 $\cdot$ 10$^4$ K) is somewhat
arbitrary and it has been taken in analogy with what done in some of
our previous publications (see in particular Recchi et al.
\cite{recc06}).  A different and maybe better choice could be 4
$\cdot$ 10$^4$ K (the minimum of the cooling curve for low values of
the metallicity -- see B\"ohringer \& Hensler \cite{bh89}, their fig.
1), or even 10$^5$ K.  We have checked how much are our results
affected by the choice of the temperature threshold.  It turns out
that the abundances in the cold phase are almost unaffected by this
choice (for $T=$ 4 $\cdot$ 10$^4$ K the abundances change negligibly
whereas for $T=$ $\cdot$ 10$^5$ K the differences are 0.02 dex at
most).  This is due to the fact that a small fraction (in mass) of gas
has temperatures in the range [2 $\cdot$ 10$^4$, 4 $\cdot$ 10$^4$] K
or [2 $\cdot$ 10$^4$, 10$^5$] K.  On the other hand, the variations in
the hot phase abundances are more significant.  They increase by up to
0.15--0.2 dex if the threshold is 10$^5$ K.  However, the overall
correlations shown in Figs. \ref{fig:oh2} and \ref{fig:ej_hotoh}
remain.

We must outline once again that the main aim of this paper was to show
the effect that the gas distribution can have on the development of
galactic winds and, hence, on the fate of gas and freshly produced
metals.  We have not attempted to reproduce the observed M-Z relation
(we will focus on this task in a paper in preparation).  For instance,
all the models, irrespective of their masses, have the same SFRs per
unit available gas mass (hence the same SF efficiencies).  Models with
initially 10$^8$ M$_\odot$ of baryonic mass have SFRs ten times larger
than models with $M_b=10^7$ M$_\odot$.  The metal production rates are
also ten times larger.  However, these larger fractions of metals are
mixed with ten times more pristine gas.  Hence, if galactic winds do
not play a significant role, the final metallicities of the models
should be approximately independent on the galactic mass (see roundish
models in Fig. \ref{fig:oh}).  However, it is known from chemical
evolution studies (e.g. Matteucci \cite{matt94}; Pipino \& Matteucci
\cite{pm04}) that SF efficiencies must increase with galactic masses
in order to reproduce the observed chemical abundances and abundance
ratios of galaxies.  This is also consistent with the idea of having
SF efficiencies increasing with pressure of the ambient diffuse gas,
which increases with galactic mass (Elmegreen \& Efremov \cite{ee97};
Harfst et al. \cite{harf06}; see also Leroy et al. \cite{leroy08}).

However, we have noticed in Sect.  \ref{subs:SFH} that metal ejection
efficiencies of galaxies with higher SFRs (but producing at the end of
the simulation the same amount of stars) are larger.  Consequently,
these galaxies tend to have lower metallicities during the SF episode.
This qualitative trend is in agreement with the so-called {\it
  fundamental metallicity relation} (FMR: Mannucci et al.
\cite{mann10}), according to which the oxygen abundance in star
forming galaxies correlates with the quantity $\mu_\alpha = \log
(M_{*}) - \alpha \log (SFR)$, where $\alpha$ is a free parameter
chosen to minimize the scatter in the FMR.  Although the precise shape
of the FMR is debated (for instance the recent paper of Andrews \&
Martini \cite{am12} find $\alpha=0.66$, more than twice the value of
Mannucci et al.), undoubtedly the SFR plays a very important role in
determining the gas-phase metallicity of a galaxy.  Galaxies with the
same stellar mass but with higher SFRs are characterized by lower
metallicities (see e.g. fig. 11 of Andrews \& Martini \cite{am12}).
According to Andrews \& Martini (\cite{am12}), this is probably due to
the fact that galaxies with high SFRs are presently experiencing a
merging.  Major mergers drive in considerable amounts of low
metallicity gas from large radii, which dilutes the metallicity of the
galaxy and triggers a vigorous SF burst.  Since most of the galaxies
with large SFR have already large stellar masses, it seems unlikely
that their gas components have still a low metallicity.  It seems to
us more likely that large SFRs drive large outflows, with a consequent
significant metal loss, as our models show.

What we want to point out here is that, for models with the same mass
and SFR, geometry plays a very significant role in determining the
fate of freshly produced metals and, consequently, the final
metallicity.  An inspection of Table \ref{table:2} shows that, for
some models, ejection efficiencies of metals can change by up to an
order of magnitude depending on the degree of flattening, being
instead the gas ejection efficiencies quite independent on $b/a$.
Consequently, the final metallicities of models with the same mass can
vary up to 1 dex depending on the geometry (see Figs.  \ref{fig:oh}
and \ref{fig:oh2}).  The large spread observed in the metallicity of
DGs with the same masses (see e.g. Lee et al.  \cite{lee06}, their
fig. 8) could be due to the effect of gas distribution.  Although the
parameter $\alpha$ is chosen to minimize the scatter in the FMR, still
galaxies with the same stellar mass and SFR show a spread in the metal
content (see e.g.  figs.  11 and 12 of Andrews \& Martini
\cite{am12}).  The gas distribution might be responsible for this
spread.

A last comment related to the M-Z relation (or to the FMR) concerns
the correlation between mass and degree of flattening of DGs.  Recent
studies suggest that smaller DGs tend to have larger axial ratios
(i.e. they tend to be rounder) than DGs with larger masses.  This
might complicate the interpretation of the M-Z and FMR relations
according to our models.  However, this effect seems to be quite
limited.  For instance, S{\'a}nchez-Janssen et al.  (\cite{sj10}) show
(their fig.  1) that the average axial ratios change by at most $\sim$
20\% between DGs with stellar masses of 10$^7$ and 10$^9$ M$_\odot$,
and that the spread in axial ratios is extremely large at all stellar
masses.  Moreover, Lisker et al.  (\cite{lisk09}) show that (at least
for DGs in clusters) the axial ratio is more related to the DG
velocity than to the mass.

Our main conclusions can be summarized as follows:
\begin{itemize}
\item The gas distribution in a galaxy plays a very important role in
  determining the fate of freshly produced metals.  Model galaxies
  with the same masses but with different degrees of flattening can
  have metal ejection efficiencies differing for one order of
  magnitude.  In particular, flat galaxies easily develop bipolar 
  outflows, through which a large fraction of metals can be lost.
\item On the other hand, the fate of pristine gas is less dependent on
  geometry than the fate of metals.  Since some models (with the same
  initial mass) can have similar ejection efficiencies of pristine gas
  but very different ejection efficiencies of metals, the final
  attained metallicities can vary by up to one dex (being the roundish
  models characterized by larger final metallicities).
\item Models characterized by the same degree of flattening show a
  clear dependence of the metal ejection efficiencies on the galactic
  mass.  Smaller galaxies (with shallower potential wells) more easily
  develop large-scale outflows, therefore the fraction of lost metals 
  tends to be higher.
\item Ejection efficiencies (of gas and freshly produced metals)
  significantly depend also on the star formation history of the
  galaxy and on the presence (or absence) of density perturbations.
\item The ejection efficiencies show instead a moderate dependence on
  the initial metallicity and on the size of the region in which the
  energetic and chemical feedback from Supernovae and stellar winds is
  redistributed.
\end{itemize}

\begin{acknowledgements}
  We thank the anonymous referee for very helpful comments and
  remarks.
\end{acknowledgements}


\begin{thebibliography}{}



\bibitem[2012]{am12} Andrews, B.~H., \& Martini, P.\ 2012, arXiv:1211.3418 

\bibitem[2002]{dm02} de Avillez, M.~A., \& Mac Low, M.-M.\ 2002, ApJ,
  581, 1047

\bibitem[1986]{bd86} Bedogni, R., \& Dercole, A.\ 1986, A\&A, 157, 101 

\bibitem[1989]{bh89} Boehringer, H., \& Hensler, G.\ 1989, A\&A, 215, 147 

\bibitem[1977]{cm77} Cowie, L.~L., \& McKee, C.~F.\ 1977, ApJ, 211, 135 

\bibitem[1986]{ds86} Dekel, A., \& Silk, J.\ 1986, ApJ, 303, 39 

\bibitem[1999]{db99} D'Ercole, A., \& Brighenti, F.\ 1999, MNRAS, 309, 941 


\bibitem[1997]{ee97} Elmegreen, B.~G., \& Efremov, Y.~N.\ 1997, ApJ,
  480, 235

\bibitem[2006]{erb06} Erb, D.~K., Shapley, A.~E., Pettini, M., et al.\
  2006, ApJ, 644, 813

\bibitem[2000]{ft00} Ferrara, A., \& Tolstoy, E.\ 2000, MNRAS, 313, 291 

\bibitem[2004]{frag04} Fragile, P.~C., Murray, S.~D., \& Lin,
  D.~N.~C.\ 2004, ApJ, 617, 1077

\bibitem[2006]{harf06} Harfst, S., Theis, C., \& Hensler, G.\ 2006,
  A\&A, 449, 509

\bibitem[1998]{hens98} Hensler, G., Dickow, R., Junkes, N., \&
  Gallagher, J.~S.\ 1998, ApJ, 502, L17

\bibitem[2004]{hens04} Hensler, G., Theis, C., \& Gallagher, J.~S.,
  III.\ 2004, A\&A, 426, 25

\bibitem[2011]{hida11} Hidalgo, S.~L., Aparicio, A., Skillman, E., et
  al.\ 2011, ApJ, 730, 14

\bibitem[2012]{hopk12} Hopkins, P.~F., Quataert, E., \& Murray, N.\
  2012, MNRAS, 421, 3522

\bibitem[2008]{kirby08} Kirby, E.~N., Simon, J.~D., Geha, M.,
  Guhathakurta, P., \& Frebel, A.\ 2008, ApJ, 685, L43

\bibitem[1974]{lars74} Larson, R.~B.\ 1974, MNRAS, 169, 229

\bibitem[2011]{lask11} Laskar, T., Berger, E., \& Chary, R.-R.\ 2011,
  ApJ, 739, 1

\bibitem[2006]{lee06} Lee, H., Skillman, E.~D., Cannon, J.~M., et al.\
  2006, ApJ, 647, 970

\bibitem[2008]{leroy08} Leroy, A.~K., Walter, F., Brinks, E., et al.\
  2008, AJ, 136, 2782

\bibitem[2009]{lisk09} Lisker, T., Janz, J., Hensler, G., et al.\
  2009, ApJ, 706, L124

\bibitem[1999]{mf99} Mac Low, M.-M., \& Ferrara, A.\ 1999, ApJ, 513,
  142 (MF99)

\bibitem[2010a]{mcqu10_1} McQuinn, K.~B.~W., Skillman, E.~D., Cannon,
  J.~M., et al.\ 2010, ApJ, 721, 297

\bibitem[2010b]{mcqu10_2} McQuinn, K.~B.~W., Skillman, E.~D., Cannon,
  J.~M., et al.\ 2010, ApJ, 724, 49

\bibitem[2001]{mfr01} Madau, P., Ferrara, A., \& Rees, M.~J.\ 2001,
  ApJ, 555, 92

\bibitem[2008]{maio08} Maiolino, R., Nagao, T., Grazian, A., et al.\
  2008, A\&A, 488, 463

\bibitem[2010]{mann10} Mannucci, F., Cresci, G., Maiolino, R.,
  Marconi, A., \& Gnerucci, A.\ 2010, MNRAS, 408, 2115

\bibitem[2011]{mann11} Mannucci, F., Salvaterra, R., \& Campisi,
  M.~A.\ 2011, MNRAS, 414, 1263

\bibitem[2006]{marc06} Marcolini, A., D'Ercole, A., Brighenti, F., \&
  Recchi, S.\ 2006, MNRAS, 371, 643

\bibitem[2002]{mart02} Martin, C.~L., Kobulnicky, H.~A., \& Heckman,
  T.~M.\ 2002, ApJ, 574, 663

\bibitem[1994]{matt94} Matteucci, F.\ 1994, A\&A, 288, 57 

\bibitem[2007]{mich07} Michielsen, D., Valcke, S., \& de Rijcke, S.\
  2007, EAS Publications Series, 24, 287

\bibitem[2010a]{mone10_1} Monelli, M., Gallart, C., Hidalgo, S.~L., et
  al.\ 2010, ApJ, 722, 1864

\bibitem[2010b]{mone10_2} Monelli, M., Hidalgo, S.~L., Stetson, P.~B.,
  et al.\ 2010, ApJ, 720, 1225

\bibitem[2002]{mfm02} Mori, M., Ferrara, A., \& Madau, P.\ 2002, ApJ,
  571, 40

\bibitem[2012]{newm12} Newman, S.~F., Genzel, R.,
  F{\"o}rster-Schreiber, N.~M., et al.\ 2012, ApJ, 761, 43

\bibitem[2005]{owb05} Ott, J., Walter, F., \& Brinks, E.\ 2005,
  MNRAS, 358, 1423

\bibitem[2012]{pan12} Pan, L., Desch, S.~J., Scannapieco, E., \&
  Timmes, F.~X.\ 2012, ApJ, 756, 102

\bibitem[1996]{pss96} Persic, M., Salucci, P., \& Stel, F.\ 1996,
  MNRAS, 281, 27

\bibitem[2004]{pm04} Pipino, A., \& Matteucci, F.\ 2004, MNRAS, 347,
  968

\bibitem[2006]{rh06} Recchi, S., \& Hensler, G.\ 2006, A\&A, 445, L39 

\bibitem[2007]{rh07} Recchi, S., \& Hensler, G.\ 2007, A\&A, 476, 841
  (RH07)

\bibitem[2009]{rha09} Recchi, S., Hensler, G., \& Anelli, D.\ 2009,
  arXiv:0901.1976

\bibitem[2006]{recc06} Recchi, S., Hensler, G., Angeretti, L., \&
  Matteucci, F.\ 2006, A\&A, 445, 875

\bibitem[2001]{rmd01} Recchi, S., Matteucci, F., \& D'Ercole, A.\
  2001, MNRAS, 322, 800

\bibitem[2002]{recc02} Recchi, S., Matteucci, F., D'Ercole, A., \&
  Tosi, M.\ 2002, A\&A, 384, 799

\bibitem[2004]{recc04} Recchi, S., Matteucci, F., D'Ercole, A., \&
  Tosi, M.\ 2004, A\&A, 426, 37

\bibitem[2007]{recc07} Recchi, S., Theis, C., Kroupa, P., \& Hensler,
  G.\ 2007, A\&A, 470, L5

\bibitem[2003]{rh03} Rieschick, A., \& Hensler, G.\ 2003, Ap\&SS, 284, 861 

\bibitem[2011]{rodr11} Rodr{\'{\i}}guez-Gonz{\'a}lez, A., Esquivel,
  A., Raga, A.~C., \& Col{\'{\i}}n, P.\ 2011, Revista Mexicana de
  Astronomia y Astrofisica Conference Series, 40, 86

\bibitem[2010]{sj10} S{\'a}nchez-Janssen, R., M{\'e}ndez-Abreu, J., \&
  Aguerri, J.~A.~L.\ 2010, MNRAS, 406, L65

\bibitem[2010]{sb10} Scannapieco, E., \& Br{\"u}ggen, M.\ 2010,
  MNRAS, 405, 1634

\bibitem[2011]{schr11} Schroyen, J., de Rijcke, S., Valcke, S.,
  Cloet-Osselaer, A., \& Dejonghe, H.\ 2011, MNRAS, 416, 601

\bibitem[1998]{stt98} Silich, S.~A., \& Tenorio-Tagle, G.\ 1998,
  MNRAS, 299, 249

\bibitem[2001]{stt01} Silich, S., \& Tenorio-Tagle, G.\ 2001, ApJ,
  552, 91

\bibitem[1989]{skh89} Skillman, E.~D., Kennicutt, R.~C., \& Hodge,
  P.~W.\ 1989, ApJ, 347, 875

\bibitem[1956]{spit56} Spitzer, L.\ 1956, Physics of Fully Ionized
  Gases, New York: Interscience Publishers

\bibitem[1953]{sh53} Spitzer, L., \& H\"arm, R.\ 1953, Phys. Rev., 89,
  977

\bibitem[2000]{ss00} Strickland, D.~K., \& Stevens, I.~R.\ 2000,
  MNRAS, 314, 511

\bibitem[1996]{tt96} Tenorio-Tagle, G.\ 1996, AJ, 111, 1641

\bibitem[2007]{tt07} Tenorio-Tagle, G., W{\"u}nsch, R., Silich, S., \&
  Palou{\v s}, J.\ 2007, ApJ, 658, 1196

\bibitem[2009]{teschio09} Tescari, E., Viel, M., Tornatore, L., \&
  Borgani, S.\ 2009, MNRAS, 397, 411

\bibitem[2004]{trem04} Tremonti, C.~A., Heckman, T.~M., Kauffmann, G.,
  et al.\ 2004, ApJ, 613, 898

\bibitem[1986]{vader86} Vader, J.~P.\ 1986, ApJ, 305, 669

\bibitem[2008]{vvs08} Vasiliev, E.~O., Vorobyov, E.~I., \& Shchekinov,
  Y.~A.\ 2008, A\&A, 489, 505

\bibitem[2012]{voro12} Vorobyov, E.~I., Recchi, S., \& Hensler, G.\
  2012, A\&A, 543, A129

\bibitem[2012]{wuyts12} Wuyts, E., Rigby, J.~R., Sharon, K., \&
  Gladders, M.~D.\ 2012, ApJ, 755, 73

\bibitem[2012]{yk12} Yang, C.-C., \& Krumholz, M.\ 2012, ApJ, 758, 48 

\bibitem[2010]{zhao10} Zhao, Y., Gao, Y., \& Gu, Q.\ 2010, ApJ, 710,
  663






\end{thebibliography}
\end{document}